\documentstyle[12pt, aaspp4]{article}
 
\begin{document}

\title{A large stellar evolution database for population
synthesis studies. II. Stellar models and isochrones for an $\alpha-$enhanced
metal distribution}
 
\author{Adriano Pietrinferni\altaffilmark{1}}
\affil{Osservatorio Astronomico di Teramo, Via M. Maggini,
64100 Teramo, Italy; pietrinferni@oa-teramo.inaf.it}
  
\author{Santi Cassisi}
\affil{INAF - Osservatorio Astronomico di Teramo, Via M. Maggini,
64100 Teramo, Italy; cassisi@oa-teramo.inaf.it}
 
\author{Maurizio Salaris}
\affil{Astrophysics Research Institute, Liverpool John Moores University,
Twelve Quays House, Birkenhead, CH41 1LD, UK; ms@astro.livjm.ac.uk}

\and

\author{Fiorella Castelli\altaffilmark{2}}
\affil{INAF - Osservatorio Astronomico di Trieste, via Tiepolo 11, 34131 
Trieste, Italy; castelli@ts.astro.it}
 
\altaffiltext{1}{Dipartimento di Statistica, Universit\`a di Teramo,
Viale F. Crucioli, 64100 Teramo, Italy}
\altaffiltext{2}{INAF-Istituto di Astrofisica Spaziale e Fisica Cosmica, Via 
del Fosso del Cavaliere, 00133 Roma, Italy}
 
\normalsize

\begin{abstract}
 
\noindent
We present a large, new set of stellar evolution models and
isochrones for an $\alpha$-enhanced metal distribution typical of
Galactic halo and bulge stars; it
represents a homogeneous extension of our stellar model library for a scaled-solar metal
distribution already presented in Pietrinferni et al.~(2004). 
The effect of the $\alpha-$element enhancement has been properly taken into account in
the nuclear network, opacity, equation of state and, for the first
time, the bolometric corrections, and color transformations.
This allows us to avoid the inconsistent use - common
to all $\alpha$-enhanced model libraries 
currently available - of scaled-solar 
bolometric corrections and color transformations for $\alpha$-enhanced models
and isochrones. We show how bolometric corrections to magnitudes
obtained for the $U,B$ portion of stellar spectra (i.e. not only the
Johnson-Cousins filters but also Str\"omgren $u,b$ magnitudes) for $T_{eff}\le6500$~K,
are significantly affected by the metal mixture,  
especially at the higher metallicities. 

Our models cover both an extended mass range (between $0.5M_\odot$ to
$10M_\odot$, with a fine mass spacing) and a broad metallicity
range, including 11 values of the metal mass fraction $Z$, corresponding to the range 
$-2.6\le[Fe/H]\le0.05$. The initial He mass fraction is 
$Y$=0.245 for the most metal-poor models and increases with $Z$
according to $\Delta Y/\Delta Z$=1.4, to reproduce the
initial solar He obtained from the calibration of the solar model.
Models with and without the inclusion of overshoot from the convective cores during
the central H burning phase are provided, as well as models with
different mass loss efficiencies.

We also provide complete sets of evolutionary models
for low-mass, He-burning stellar structures covering the whole metallicity range, to
enable synthetic horizontal branch simulations.

We compare our database with several widely used stellar model
libraries from different authors, as well as with various observed color
magnitude and color-color diagrams (Johnson-Cousins $BVI$ and near infrared magnitudes, 
Str\"omgren colors) of Galactic field stars and globular clusters. We also test our
isochrones comparing integrated optical colors and Surface Brightness 
Fluctuation magnitudes with selected globular cluster data. 
We find a general satisfactory agreement with the empirical constraints. 

This database, used
in combination with our scaled-solar model library, is a valuable tool 
for investigating both Galactic and extra-galactic simple and composite stellar populations,
using stellar population synthesis
techniques. 

\end{abstract}
 

\section{Introduction}

\noindent

Large libraries of stellar models and isochrones covering wide ranges
of age and metallicities are an essential tool to investigate the
properties of resolved and unresolved stellar populations. 
We have recently published (Pietrinferni et al.~2004, hereafter
Paper I) an extended database of scaled-solar 
stellar evolution models and isochrones that is very well suited to accomplish this goal.  

It is however well known that the scaled-solar metal mixture is not
universal. In particular, observations of metal abundance ratios in
the Galaxy (i.e., Ryan, Norris \& Bessel~1991; Carney~1996; Gratton, Sneden \& Carretta~2004;
Sneden~2004;) have shown that [$\alpha$/Fe] 
(where with $\alpha$ we denote collectively the so called
$\alpha$-elements  O, Ne, Mg, Si, S, Ar, Ca and Ti) is larger than
zero in Population~II field stars, Galactic globular clusters (see, e.g.,
Carney~1996) and in Galactic bulge stars (see, e.g., McWilliam \& Rich
2004) that is, the abundance ratio $\alpha$/Fe is larger than the solar counterpart. 
Typical values of [$\alpha$/Fe] in the Galactic halo are $\sim0.3 -
0.4$ dex. 
In addition, also stellar populations in elliptical galaxies appear to have formed
with a metal mixture characterized by [$\alpha$/Fe]$>$0 (see, i.e.,
Worthey, Faber \& Gonzalez~1992; Tantalo, Chiosi \& Bressan, 1998; Trager et al.~2000)

From a theoretical point of view, the first extended set of isochrones
accounting for the enhancement of oxygen was published by Bergbusch
\& Vandenberg~(1992) (and oxygen-enhanced Horizontal Branch models by
Dorman~1992); the enhancement of all $\alpha$-element
was exhaustively addressed by Salaris, Chieffi \&
Straniero~(1993) in the regime of low mass, metal poor objects, while Weiss,
Peletier \& Matteucci~(1995) studied the high metallicity regime. 
An important and useful property of $\alpha$-enhanced models 
shown by Salaris et al.~(1993) is that, for $\alpha$-enhanced mixtures that
satisfy the constraint [(C+N+O+Ne)/(Mg+Si+S+Ca+Fe)]=0 (as for example
in case of a constant enhancement of all $\alpha$ elements) low mass
stellar models computed
with an [$\alpha$/Fe]$>$0 metal distribution are equivalent to scaled-solar ones 
with the same global metal content $Z$ 
(see also Chaboyer, Sarajedini \& Demarque~1992). 
This property breaks down at values of Z larger than about Z=0.002 
(see, e.g., Weiss et al.~1995, VandenBerg \& Irwin~1997, 
Salaris \& Weiss~1998, Salasnich et al.~2000, 
VandenBerg et al.~2000, Kim et al.~2002).

These theoretical results have been obtained neglecting the effect
of an $\alpha$-enhancement on the bolometric corrections to be applied to
the theoretical isochrones. Very recently, Cassisi et al.~(2004)
addressed this issue for the first time, showing that broadband colors involving the
$UB$ part of the spectrum are affected by the enhancement of the
$\alpha$-element, whereas redder wavelength bands are not affected. 
Both $(B-V)$ and $(U-B)$ colors appear bluer than the 
scaled-solar ones at either the same [M/H]
or the same [Fe/H], the difference increases with increasing metal
content and/or decreasing 
effective temperature. The best way
to mimic $\alpha$-enhanced color transformations is to use scaled-solar 
ones with the same [Fe/H], but this equivalence breaks down when
the iron content increases above [Fe/H]$\sim -$1.6. 
These differences are mainly due to the enhancement of Mg with 
lower contributions from the enhancement of Si and O.

The use of scaled-solar isochrones to mimic their
$\alpha$-enhanced counterpart is not valid over the full metallicity
range where ratios [$\alpha$/Fe]$>$0 have been observed, so that it is
important to compute models and isochrones that 
include appropriate values of [$\alpha$/Fe].
In the last years Salaris \& Weiss~(1998), VandenBerg et al.
(2000), VandenBerg~(2000), Salasnich et al.~(2000), Kim et al.~(2002)
have published sets of $\alpha$-enhanced isochrones that cover to
different degrees the age and metallicity parameter range.
None of these sets adopt appropriate $\alpha$-enhanced bolometric
corrections and color transformations.

In this paper we enlarge the parameter space covered by the model and
isochrone library presented in Paper~I, by extending our computations
to an $\alpha$-enhanced metal distribution, where the [$\alpha$/Fe]
value is consistent with observations of the Galactic halo population. 
This new library covers
exactly the same $Y, Z$ space, and the same mass ranges of our
scaled-solar one presented in Paper~I, and it has been computed with
the same stellar evolution code and homogeneous physical inputs. It 
includes models computed both with and without overshooting from the
convective cores, two different choices for the efficiency of mass loss and, for the first time, 
appropriate $\alpha$-enhanced
color transformations and bolometric corrections for both the low-
and high $Z$ regimes. 

The paper is organized as follows: \S~2 briefly summarizes 
the adopted physical inputs, while the model library is
presented in \S~3. Comparisons with widely used isochrone
databases, and with selected empirical constraints are discussed in
\S~4, and \S~5, respectively. A summary and final remarks follow in \S~6.

\section{Input physics and color transformations}
 
\noindent
We used the same stellar evolution code as in Paper I, and the reader
is referred to this paper for more information. 
Here we just describe the new chemical and physical inputs necessary for the
computation of the $\alpha$-enhanced library.
The adopted $\alpha$-enhanced mixture is the same as in the models by Salaris,
Degl'Innocenti \& Weiss~(1997) and Salaris \& Weiss~(1998) and it is
reported in Table~1. The $\alpha$-elements have been enhanced with
respect to the Grevesse \& Noels~(1993) solar metal distribution (that
is the heavy element distribution used in Paper~I) by variable
factors, following mainly the results by Ryan et al.~(1991) about field
Population~II stars. The overall average enhancement is [$\alpha$/Fe]$\sim$0.4.
This $\alpha$-enhanced distribution has been used in the nuclear
network, the radiative (Iglesias \& Rogers~1996, Alexander \& Ferguson~1994) 
and conductive opacities (Potekhin et al.~1999) and the equation 
of state\footnote{We have employed the equation of state by
A. Irwin. His code is made publicly available at 
ftp://astroftp.phys.uvic.ca under
the GNU General Public License (GPL). A full
description of this EOS can be found at the following URL site: 
http://freeeos.sourceforge.net}.

Although our stellar evolution code can 
account for the atomic diffusion of helium 
and heavy elements, we decided to neglect this process in our model
grid computation. 
As largely discussed in Paper~I, we have accounted for atomic diffusion
when computing the standard solar models. This choice allowed us to properly calibrate
the mixing length (see below) and to estimate the initial solar
chemical composition. Although in the Sun
atomic diffusion is essentially fully efficient, spectroscopic
observations of stars in Galactic globular clusters or field halo stars 
(see section II of Paper~I) point to a drastically reduced efficiency
of this process, due possibly to the counteracting effect of rotationally
induced mixing processes. This
evidence raises the possibility that the almost uninhibited efficiency of
diffusion found in the Sun might not be an occurrence common to all stars.

Superadiabatic convection is treated according to the Cox \&
Giuli~(1968) formalism of the mixing length theory 
(B\"ohm-Vitense~1958) and 
the mixing length value is, as in the case of the scaled-solar 
models, equal to 1.913, as obtained from a calibration
of the solar model (see Paper~I for details).

All models include mass loss using the Reimers formula (Reimers 1975) with the free
parameter $\eta$ set to two different values: 0.2 and 0.4\footnote{In
the scaled-solar library presented in Paper~I we used only one value for the
parameter $\eta$, namely $\eta=0.4$. However, later we decided to
recompute the model library using also
the value $\eta=0.2$. As for all other models, 
these additional computations can be found at the URL site:
http://www.te.astro.it/BASTI/index.php}.

We made the same assumptions as in Paper~I regarding 
the treatment of convection at the border of convective core
during the core H-burning phase of stars with total mass larger than $\sim
1.1M_{\odot}$. Two different values are adopted for the overshoot from
the Schwarzschild convective boundary, namely 
$\lambda_{OV}=0$ and 0.2. With $\lambda_{OV}$ we denote the length - expressed as a fraction of
the local pressure scale height - crossed by the convective 
cells in the stable region outside the Schwarzschild boundary.
The amount of core overshoot is decreased in the same way as in
Paper~I, when the convective core size decreases to zero as a consequence of
the decreasing total stellar mass.
The treatment of convection in the core of central He-burning stars is exactly the same of Paper~I.
The possible occurrence of overshooting from the bottom of the convective envelopes has been
neglected as in Paper~I for the reasons discussed in that paper.

Finally, the theoretical tracks and isochrones have been transformed to
various observational Color-Magnitude Diagrams (CMDs) using 
color-$T_{eff}$ transformations and bolometric
corrections (that we will denote collectively as CT) obtained from an updated version
of ATLAS9 model atmospheres (Castelli \& Kurucz~2003). The scaled-solar and [$\alpha$/Fe]=0.4 CT have been
described in Paper~I and in Cassisi et al.~(2004). In the latter paper a
subset of the $\alpha$-enhanced CT has been used to 
investigate the impact of an $[\alpha/Fe]>0$ metal distribution on broadband colors.
Grids of model atmospheres, energy distribution, color indices\footnote{They are available 
at the URL site: http://wwwuser.oat.ts.astro.it/castelli} in the 
$UBVRIJKL$, Str{\"o}mgren $uvby$ and Walraven photometric
systems have been computed for $[Fe/H]$: $-2.5$, 
$-2.0$, $-1.5$, $-1.0$, $-0.5$, $0.0$, $+0.2$ and
$+0.5$. 
For a more detailed discussion about the adopted passbands and the 
model atmosphere grid the reader is referred
to Paper~I and to Cassisi et al.~(2004). Here we only add that the Str{\"o}mgren colors
used in this paper were computed according the prescriptions given by Relyea \& Kurucz~(1978).

	
\section{The evolutionary tracks and isochrones}

\noindent

The grid of chemical compositions of our library covers 11 pairs of
$Y, Z$ values, as listed in Table~2, ranging from $Z$=0.0001 to
$Z$=0.04. They are the same
$Y, Z$ values of our scaled-solar library, although the correspondence
between $Z$ and [Fe/H] is different, due to the effect of the $\alpha$-enhancement.
We wish to note that in the scaled-solar library presented in Paper~I,
the $Z=0.0006$ metallicity was originally missing; it has been added
later, in order to improve the sampling of the low-metallicity regime.
We have used  $Y=0.245$ for the lowest metallicity, together with 
$dY/dZ\sim 1.4$ that allows to match the calibrated initial He
abundance in the Sun.

We have used the same stellar mass range and sampling as in Paper~I,
to be consistent with the scaled-solar library, and to optimize the use
of both sets of models in a stellar population synthesis code. 
For each chemical composition, we have computed the evolution of up to
41 different stellar masses, with a minimum mass equal 
to $0.50M_\odot$, and a maximum mass always equal
to $10M_\odot$\footnote{We plan to extend this mass range
towards both very low-mass and more massive structures in the near future.}.
We adopted a mass step of $\Delta{M}=0.1M_\odot$ (or lower) for masses below
$2.5M_\odot$, $\Delta{M}=0.2M_\odot$ for masses in the range
$2.5\le{M/M_\odot}\le3.0$, 
and $\Delta{M}=0.5M_\odot$ for more massive
models\footnote{The grid is slightly more coarse for the models 
including core overshooting.}. 
Models less massive than $\sim 3M_\odot$ have been computed
starting from the pre-MS phase, whereas the more massive ones
have been computed starting from a chemically homogeneous
configuration on the MS. 
All models, apart from the less massive
ones whose central H-burning evolutionary times are longer than 
the Hubble time, have been evolved until the C ignition or 
after the first few thermal pulses along the Asymptotic Giant Branch
(AGB) phase. We are already working to extend the evolution of low
mass and intermediate mass stars up to the end of the thermal pulses
along the AGB, using the synthetic AGB technique (see, e.g., Marigo,
Bressan, \& Chiosi~1996 and reference therein).

\begin{figure}        
\plotone{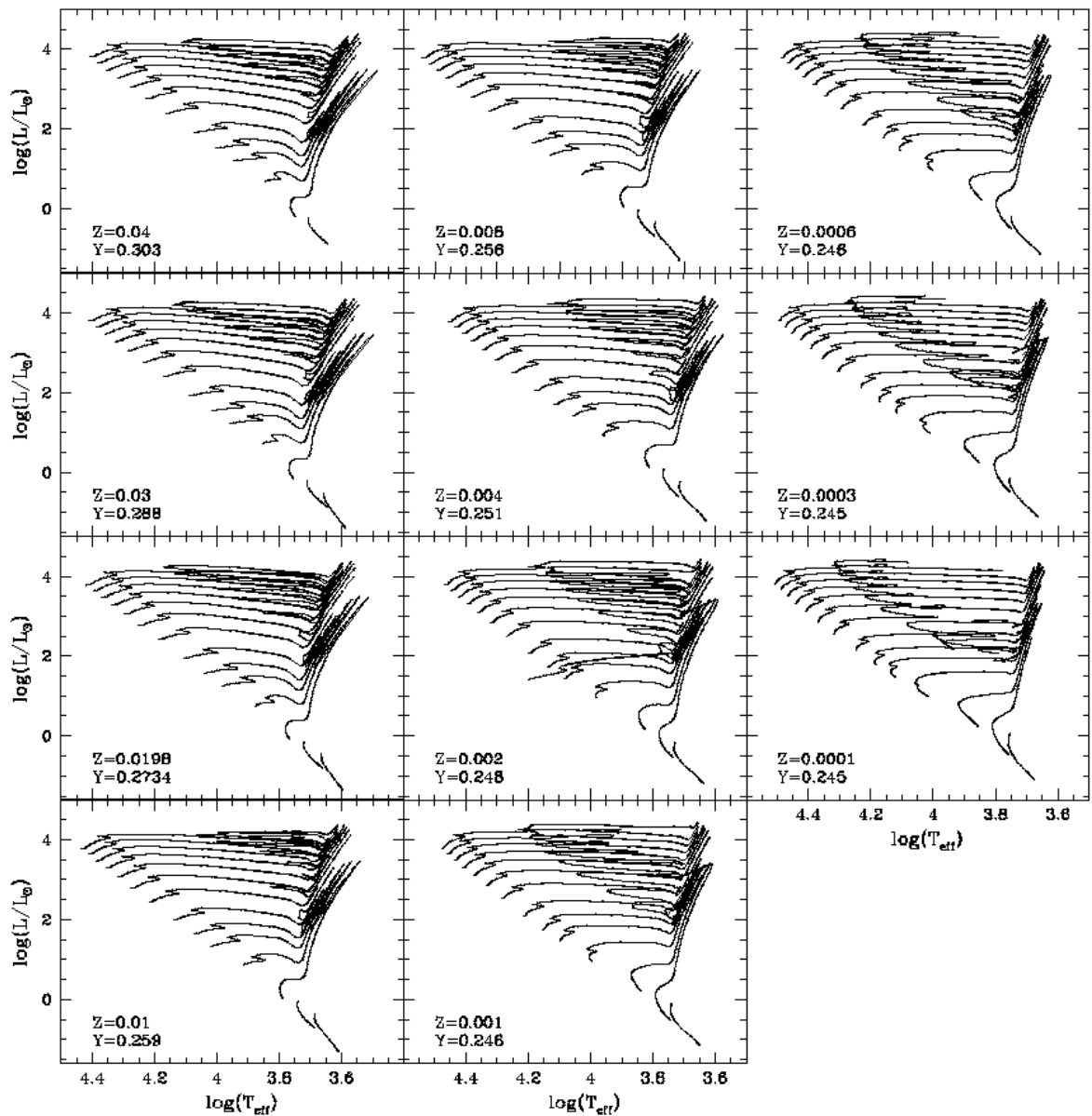}        
\caption{Selected evolutionary tracks for the 11 metallicities of our
model grid.
\label{Htracks}}        
\end{figure}

In case of models undergoing a violent He flash at the Red Giant
Branch (RGB) tip, we do
not perform a detailed numerical computation
of this phase, but we use the evolutionary values of both the He
core mass, $M_{cHe}$, and the chemical profile at the RGB tip 
to compute the corresponding Zero Age Horizontal Branch
(ZAHB) model (see Vandenberg et al.~2000 for a careful check of the validity of this approach).
A subsample of the evolutionary tracks (models without convective
overshoot and $\eta$=0.2) is shown in Fig.~\ref{Htracks} 
for the entire chemical composition range.

\begin{figure}        
\plotone{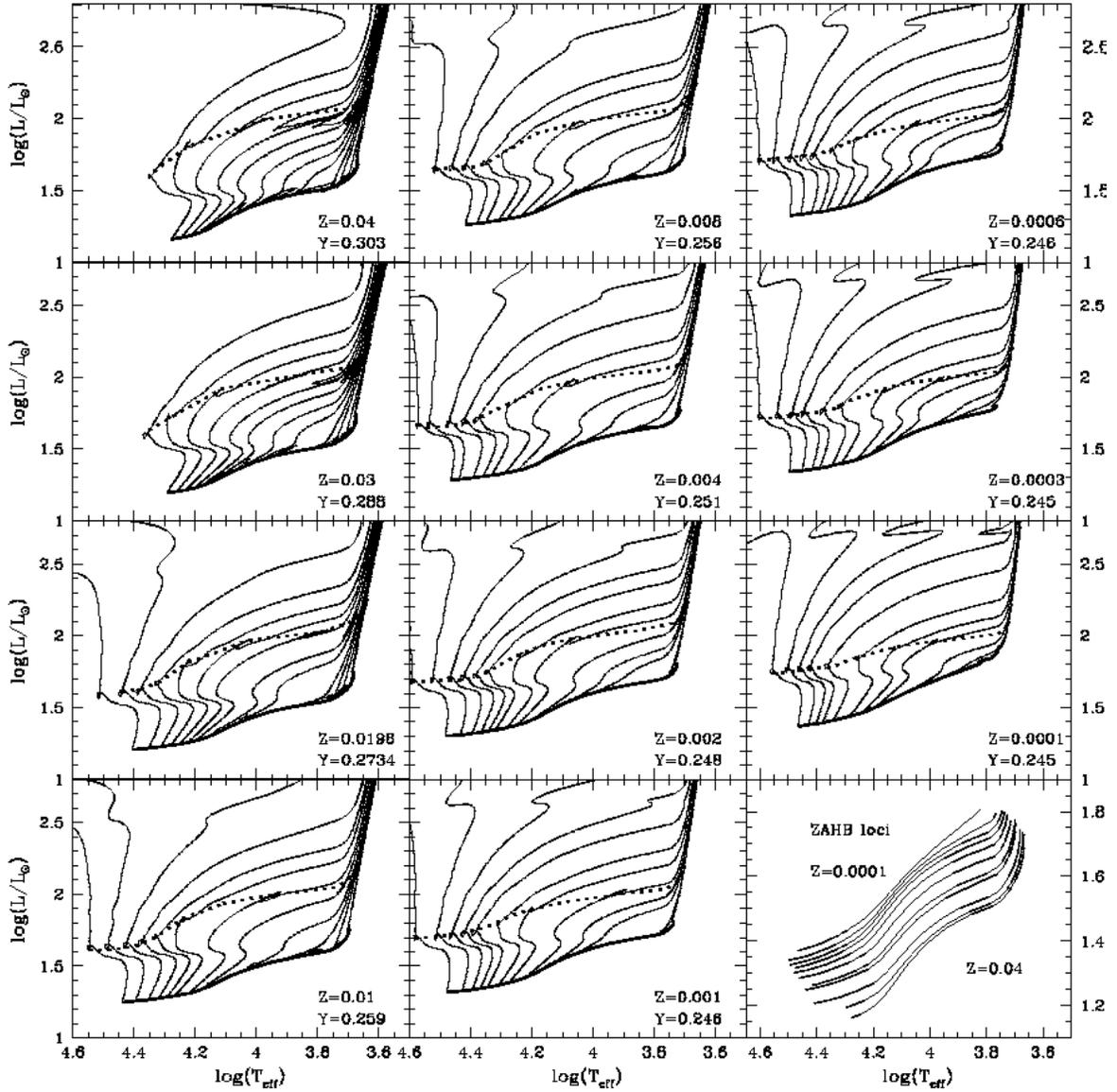}        
\caption{Additional HB evolutionary tracks suited for the computation
of synthetic HB populations. The figure shows also the location of the ZAHB and central He
exhaustion loci (dotted lines).
\label{Hetracks}}        
\end{figure}

For each chemical composition, we have also computed an extended set of He burning
models -- that is, HB models in addition to the one obtained with our
chosen mass loss law -- 
with various different values of the total stellar mass after the RGB phase,
using the fixed core mass and chemical
profiles given by an RGB progenitor having an age (at the RGB tip)
of $\sim 13$Gyr. These models allow one to compute the CMD of synthetic HB
populations, including an arbitrary spread of the amount of mass lost along the RGB
phase (typical of Galactic globular clusters).
The RGB progenitor of these additional HB models has a mass typically equal to
$0.8M_\odot$ at the lowest metallicities, increasing up 
to 1.0$M_\odot$ for the more metal-rich compositions. 
Table~\ref{zahb} lists the main characteristics
of these HB models, namely, the He core mass at the ZAHB, 
the envelope He content, the total mass, 
luminosity and absolute visual magnitude of the object  
whose ZAHB location is at $\log{T_{eff}}=3.85$ - taken as rapresentative
of the mean temperature inside the RR Lyrae instability strip.
For these additional He-burning 
computations we have employed a fine mass spacing, and more than
30 HB models have been computed for each chemical composition.
A subset of these models is displayed in Fig.~\ref{Hetracks}.
The same figure shows also the location of the ZAHB and central He-exhaustion loci.

All the post-ZAHB computations have been extended either to the onset 
of thermal pulses for more massive models, or until the luminosity 
along the white dwarf cooling sequence has decreased down to
$\log(L/L_\odot)\sim -2.5$ for less massive objects.
They allow a detailed investigation of the various evolutionary paths 
that follow the exhaustion of the central He burning.

The individual evolutionary tracks have been reduced to the
same number of points, to facilitate the computation of isochrones and
their use in a population synthesis code. Along each evolutionary
track some characteristic homologous points (key points -- KPs) 
corresponding to well-defined evolutionary phases have been identified. 
The choice and number of KPs 
had to fulfill two conditions: 1) all main evolutionary
phases have to be properly accounted for; 2) the number of KPs has to be
large enough to allow a suitable sampling of the track morphology, even for 
fast evolutionary phases. The second condition requires also that the number
of points between two consecutive KPs has to be properly chosen.

For a careful description of the adopted KPs we refer to Paper~I. 
In the original version of the database, 
we adopted 14 key stages along the evolutionary tracks (see Paper~I). 
For our actual stellar models library, we have accounted 
for 16 key stages along the evolutionary track to obtain a 
more accurate sampling of the track 
morphology (we thank A. Dolphin for pointing out to us the need of a
more accurate sampling of the RGB bump region). Of course all evolutionary tracks
for the scaled-solar metal distribution have been re-reduced by
accounting for the same 16 KPs.
The whole set of evolutionary computations (but the additional HB
models discussed above) have been used to 
compute isochrones for a large age range, namely, from
30~Myr up to 16~Gyr (older isochrones can also be computed), 
from the Zero Age Main Sequence (ZAMS) up to the first thermal pulse along the AGB 
or to the C ignition
\footnote{In common with the main large databases for population synthesis
studies, our isochrones do not include the pre-MS phase (although it 
has been computed for the individual tracks, as detailed above). For 
ages below $\approx$100~Myr, stars of $\sim 0.5~M_{\odot}$ would be
still evolving along the pre-MS phase, whereas in our isochrones they 
appear along the ZAMS. To give an idea of the systematics involved, at
intermediate metallicities and ages of 30~Myr, a
0.5$M_{\odot}$ star along our isochrones is fainter 
by $\sim$0.4~mag in $V$ and hotter by about 70~K than its real location. 
When our evolutionary computations of Very Low Mass stars will be 
completed, we will consistently include in the isochrones also the pre-MS evolution.}.

The entire library of evolutionary tracks and isochrones is 
available upon request to one of the authors or can be 
retrieved at the WEB site 
http://www.te.astro.it/BASTI/index.php\footnote{Those interested in obtaining more information,
specific evolutionary results as well as additional set of stellar models can
contact directly one of the authors or use the request form available at the quoted Web site.}.

At the same URL site, we have made available a World Wide Web interface
(more details will be provided in a forthcoming paper, 
D. Cordier et al.~2006 in preparation)
that allows a user to compute isochrones, evolutionary tracks and luminosity functions for any
specified stellar mass, chemical composition, age. Detailed tables
displaying the relevant evolutionary features
for all computed tracks are also available.

The files containing the isochrones provide at each point (2000 points
in total) the initial value of the evolving mass, the actual mass
value (in principle different, due to the effect of mass loss), 
$\log(L/L_{\odot})$, $\log{T_{eff}}$, and absolute magnitudes in
different photometric systems 
(see the discussion in the next section).

These isochrones - as well as those for a scaled-solar metal
distribution - can be used as input data for 
our fortran code $SYNTHETIC MAN(ager)$ that we have written in 
order to compute synthetic CMDs,
integrated colors and integrated magnitudes of a generic stellar
population with an arbitrarily chosen star formation rate and
age metallicity relation. 
A presentation of this code can be found in Paper~I and it will not be repeated
here. In the near future, we will allow users to run this code through
a P.E.R.L based, Web interface\footnote{A pre-release 
version can be seen at http://astro.ensc-rennes.fr/basti/synth\_pop/\
index.html}. In addition to a chosen Star Formation History the user 
will be free to fix various other 
parameters like photometric and spectroscopic errors, color excess, 
fraction of unresolved binaries, geometrical depth of the
population. The access to 
this tool will require 
registration\footnote{Pre-registration can be done by sending an e-mail to 
S. Cassisi or D. Cordier}, the results being sent to the user by e-mail.

\subsection{The impact of an $\alpha-$enhanced metal distribution on HB structures}

\noindent
Before closing this section, 
we wish to briefly discuss the effect of an $\alpha$-enhanced
metal distribution on the HB evolution. Although the effect 
of $\alpha-$enhanced mixtures on H-burning models is well known and widely
discussed in the literature (see the references in the introduction) the
behavior of $\alpha$-enhanced He-burning models compared to their scaled-solar
counterpart has been investigated only by Salaris et al.~(1993) and 
VandenBerg et al.~(2000), although the latter studied the effect
on the ZAHB location only. An earlier investigation by Bencivenni et
al.~(1989) did not take into account the effect of an
[$\alpha$/Fe]$>$0 mixture on the HB progenitor evolution.

\begin{figure}        
\plotone{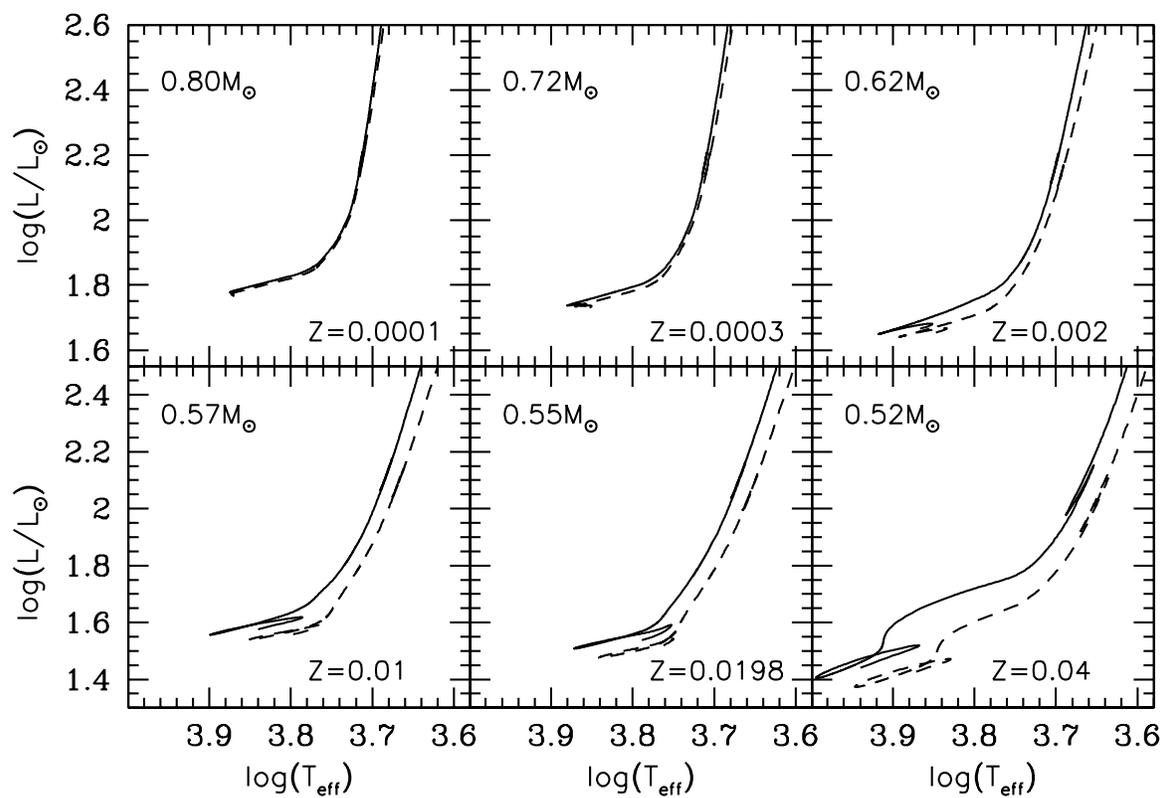}        
\caption{Comparison between an $\alpha-$enhanced (solid line) 
and scaled-solar (dashed line) HB tracks in the instability strip, 
for the labeled values of metallicity and total mass.
\label{hba}}        
\end{figure}

Figure~\ref{hba} shows the comparison between an 
$\alpha-$enhanced HB model in the instability strip and the corresponding scaled-solar one at
fixed total mass for various values of $Z$. 
At fixed total mass and global metallicity, $\alpha-$enhanced HB models
appear brighter and hotter than the scaled-solar ones, and in particular
for values of $Z$ larger than $\sim$0.002.
At $Z=10^{-4}$, the luminosity difference is negligible,
$\Delta\log(L/L_\odot)\approx0.004$, while at $Z=0.002$, $\Delta\log(L/L_\odot)\approx0.02$.
These differences increase to $\sim0.03$ and $\sim0.05$ at $Z=0.0198$ and
$Z$=0.04, respectively.
These results about the brighter and hotter location of
$\alpha$-enhanced HB models in the instability strip agree well 
with the findings by Salaris et al.~(1993) and VandenBerg et
al.~(2000). For $Z\geq$0.002 the $\alpha-$enhanced ZAHB models have a 
He core mass lower than the scaled-solar case by only $\sim 0.001M_\odot$ (regardless of the
metallicity) and an envelope
He abundance essentially equal to or lower by at most $\Delta
Y\sim$0.001; this means that the difference in the ZAHB location and
HB evolution is due to the increased efficiency
of the CNO-cycle in the H-burning shell with respect to the core burning 
in models with an [$\alpha$/Fe]$>$0 distribution.

As for the lifetime along the HB phase for a given $Z$, 
we find that it is only marginally affected by the use of an
$\alpha-$enhanced metal distribution, irrespective of the precise
value of $Z$. The central He-burning
lifetime decreases compared to the scaled-solar mixture 
by a negligible $\sim1$\% at low metallicity, and about $\sim4$\% at the
highest metallicity of our model grid. 
The same change holds for the AGB lifetime up to the thermal pulse
phase. 
This has the important consequence that the theoretical calibrations of
the $R_2$ parameter (number ratio of AGB stars to HB stars; see
e.g. Caputo et al.~1989) is not affected by the adopted metal distribution.
This parameter is very important for its sensitivity to the extension of the convective
cores during the central He-burning phase, and is used to assess
the efficiency of the breathing pulses at the exhaustion of the
central He.

A further important evolutionary feature is the brightness of the AGB
clump, 
marking the ignition of the He-burning in a shell. We have verified  
that the brightness of the AGB clump is slightly larger with respect
the scaled-solar case, the difference tracking the luminosity
difference of the ZAHB models. This implies that the 
brightness difference between the ZAHB and the AGB clump
is preserved when passing from a scaled-solar mixture to an
$\alpha-$enhanced one with the same $Z$, for any value of $Z$.

\subsection{$\alpha-$enhanced vs scaled-solar transformations}

\begin{figure}[t]
\plotone{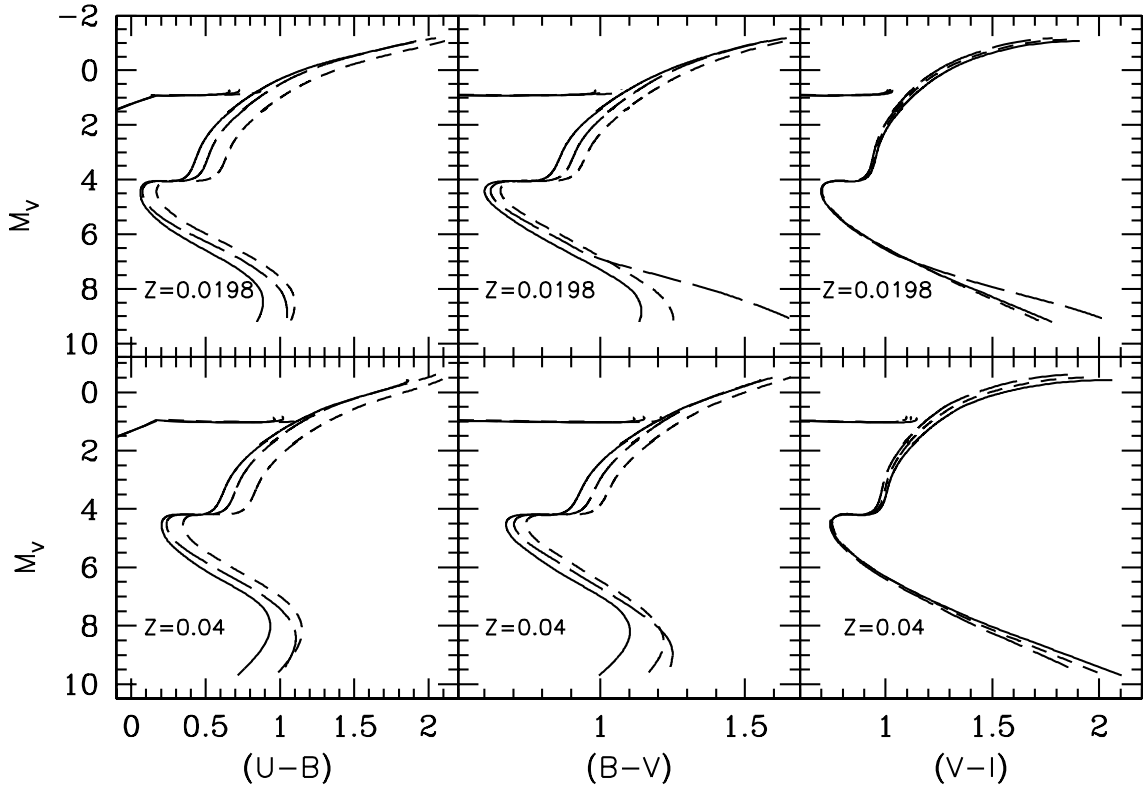}   
\caption{CMDs of 12~Gyr old $\alpha-$enhanced isochrones with the
labeled metallicities. Three different sets of CT transformations
have been employed. Solid lines represent $\alpha-$enhanced
transformations with the appropriate metal mixture; long dashed lines
denote scaled-solar transformations with the same [Fe/H] of the
$\alpha-$enhanced isochrone, whereas short dashed lines display the case of
scaled-solar transformations with the same [M/H] of the
$\alpha-$enhanced isochrone.  
\label{ubvihz}}   
\end{figure}

\noindent
Cassisi et al.~(2004) analized the impact of $\alpha$-enhanced CT in
the $UBVRIJHKL$ filters, for the metallicity regime of Galactic globular
clusters, and their results have been already summarized in the introduction.
Here we extend the analysis to higher metallicities. 
Figure~\ref{ubvihz} shows the comparison between 12~Gyr old
$\alpha$-enhanced 
isochrones and ZAHB for $Z$=0.0198 and $Z$=0.04 in various CMDs, using different CT.
More in detail, we used the appropriate $\alpha$-enhanced CT, a
scaled-solar CT with the same [Fe/H] of the isochrone, and a scaled-solar CT with
the same [M/H] (total metallicity) of the isochrone. 

We find (in agreement with the results at lower $Z$) that the $V$
magnitudes are hardly affected, whereas the $(U-B)$ and $(B-V)$
colors predicted by the scaled-solar CT with the same [M/H] of the
isochrone are systematically redder along the whole
isochrone. Differences are smaller when considering scaled-solar CT
with the same [Fe/H] of the isochrone. This occurrence means that it is 
the abundance of Fe and of all other scaled-solar elements that contribute
mostly to the observed color, although not completely, given the
difference with the appropriate CT -- but they still reach up to   
$\sim$ 0.1~mag along a large fraction of the isochrone. This is due to the
effect of the $\alpha$-enhancement in the wavelength range of the $U$
and $B$ filter (see discussion in Cassisi et al.~2004).

A different approach to mimic $\alpha$-enhanced
CT with scaled solar ones, is to use scaled-solar CT with the
same [$\alpha$/H] ratio of the [$\alpha$/Fe]$>$0 distribution. 
So doing, one assumes
that it is essentially the abundance of $\alpha$-elements 
that affects the colors. 
The data in Figure~\ref{ubvihz} show that this choice is much less accurate, since it 
corresponds to a scaled-solar CT with a [M/H] value 0.4~dex
larger that the actual one, that yields even redder colors.

Interestingly, the $(V-I)$ color is also affected by the CT
choice at high metallicities and low temperatures. 
The two scaled-solar CT produce colors this time systematically bluer than the
appropriate $\alpha$-enhanced CT, the CT with the same [Fe/H] of the
isochrone giving the worst results. On the basis of our previous
discussion it is clear that in this case a scaled-solar CT with the
same [$\alpha$/H] ratio of the isochrone would provide colors closer to
the appropriate ones. This suggests that the abundance of the $\alpha$-elements (or
at least some of them) starts to dominate the CT in this wavelength
range for solar and super-solar metallicities, and low temperatures.

\begin{figure}        
\plotone{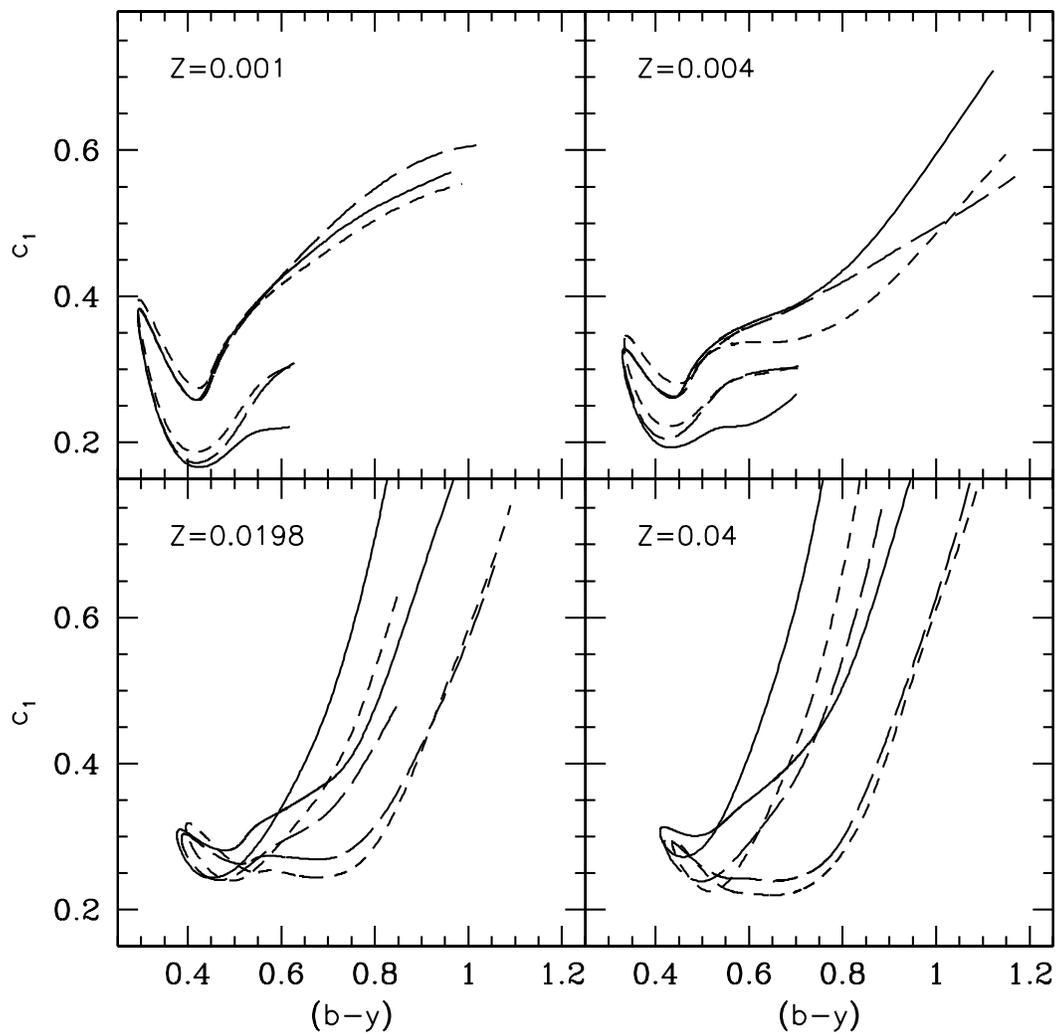}        
\caption{As fig.\ref{ubvihz}, but for Str{\"o}mgren colors.
\label{strom1}}        
\end{figure}

Figure~\ref{strom1} shows another comparison of CT applied to a 
12~Gyr old $\alpha$-enhanced isochrone at various $Z$, considering the
appropriate $\alpha$-enhanced transformations and the same two choices for
the scaled-solar CT as in Fig.~\ref{ubvihz}, this
time involving the Str\"omgren filters. The Str\"omgren $u$, $v$ and $b$
filters are located in the wavelength range of the Johnson $U$ and $B$
bands; therefore one expects a relevant influence of the
$\alpha$-elements on, for example, the  $[c_1, b-y]$ color-color
plane, used to estimate ages of star clusters and field stars, 
independently of the knowledge of their distance.

\begin{figure}        
\plotone{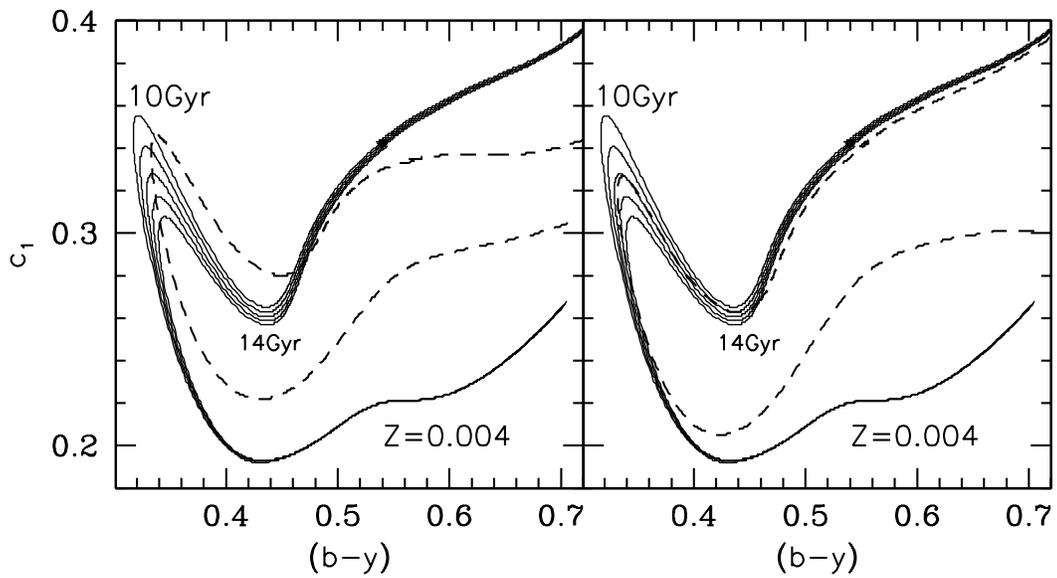}        
\caption{$\alpha-$enhanced isochrones with age in the range 10-14~Gyr at step of 1~Gyr, transformed
in the $[c_1, b-y]$ diagram by using $\alpha-$enhanced. The dashed line shows the location of one of these
isochrones, namely the 12~Gyr old one, but transformed by using alternatively a scaled-solar CT at the same global metallicity (left panel)
or at the same [Fe/H] of the $\alpha-$enhanced one (right panel).
\label{strom2}}        
\end{figure}

The use of different CT affects strongly 
the isochrone morphology in this plane, as shown clearly by Fig.~\ref{strom1}. 
The effect is appreciable from $Z\sim$0.001 and 
is, as expected, larger at larger metallicities.
The sections of the
isochrone that are more affected are, as for the case of Johnson colors, 
the RGB and the lower MS. Again, scaled-solar CT at the same [Fe/H] of
the appropriate ones give a better approximation to the real CT, apart
from the highest $Z$, where differences are huge for both scaled solar CT.
In Fig.~\ref{strom2} we show isochrones 
for various ages and $Z=0.004$, transformed to the  $[c_1, b-y]$ plane
using the same three choices of CT. 
One can notice that the 12~Gyr old isochrone
transformed using the scaled-solar CT with the appropriate [Fe/H]
matches well Turn Off (TO) and subgiant branch obtained with the appropriate
colors. On the contrary, using scaled-solar CT with the same [M/H] of the
$\alpha$-enhanced ones causes not only a change of the morphology of
the lower MS and the RGB, but also a shift of both $c_1$ and $b-y$ of
the upper MS, TO and subgiant region; 
this induces systematic errors in both the age and reddening determined
from isochrone fitting.
 
\section{Comparison with existing databases}

In this section we discuss the comparison in the H-R diagram of selected isochrones with 
similar predicitions from the databases by Salasnich et
al.~(2000), Salaris \& Weiss~(1998), VandenBerg et al.~(2000) and Kim
et al.~(2002).

These grids of
models are computed employing some different choices of the physical
inputs (the source of the opacities is quite often the same, but 
equation of state, nuclear reaction rates, boundary conditions and 
neutrino energy loss rates are often different) 
and also the chemical compositions are often different.
We perform the comparison on the theoretical H-R diagram, thus bypassing
the additional degree of freedom introduced by the choice of the
color transformations. 
In each comparison we have selected  
chemical compositions as close as possible to our choices. It is
important to notice that Kim et al~(2002) models do include He
diffusion, and therefore their isochrones for old ages (above a few Gyr) are
affected by the efficiency of this process.

\begin{figure}        
\plotone{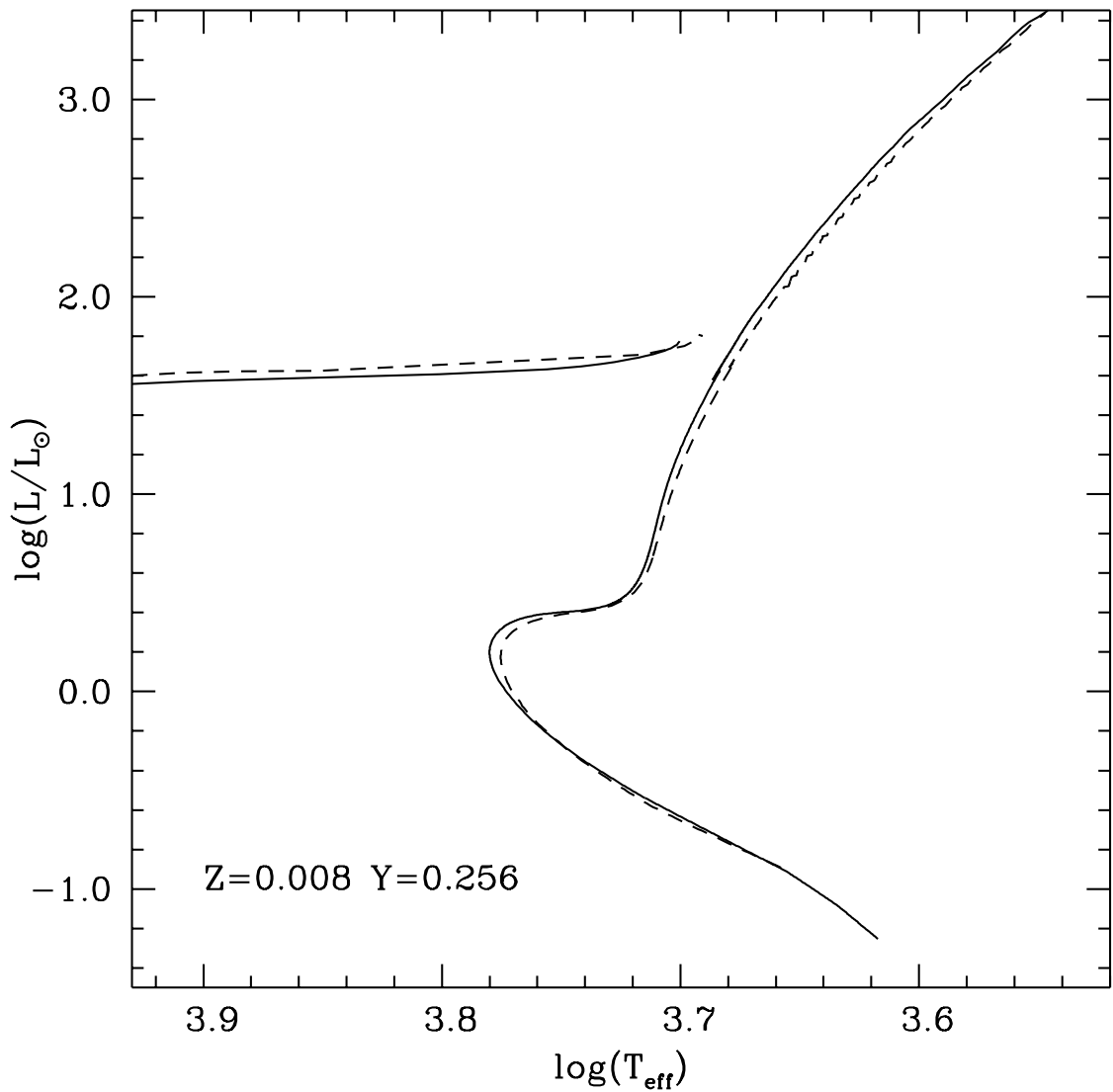}        
\caption{Comparison between our 12~Gyr old isochrone (solid line) with the
labeled chemical composition, and a 12~Gyr old, $Z=0.008$
$\alpha$-enhanced isochrone (dashed line) from Salaris \& Weiss~(1998).
\label{SW1}}        
\end{figure}

Figure~\ref{SW1} shows the comparison in the H-R diagram of a $Z=0.008$, 12~Gyr old isochrone
from Salaris \& Weiss~(1998 - hereafter SW98) with our own isochrone with the same age and 
metallicity, from the
ZAMS to the TRGB. The ZAHBs are also displayed.
Metal mixture and opacities are exactly the same in two 
models, and also the initial He mass fraction is practically identical
($Y=0.254$ in SW98 models, compared to our adopted
value of $Y=0.256$).  

There are some differences in the luminosities of the
ZAHB and to a lesser extent the TO of the two isochrones. The ZAHB
from SW98 is brighter by $\Delta\log(L/L_{\odot}) \sim$0.03-0.04, whereas the
TO is fainter by  $\Delta\log(L/L_{\odot}) \sim$0.02. Effective
temperatures along the MS are in good agreement, but the temperature at
the TO region and along large part of the RGB of our isochrones is
hotter by about 70~K and $\sim$50~K, respectively.

\begin{figure}        
\plotone{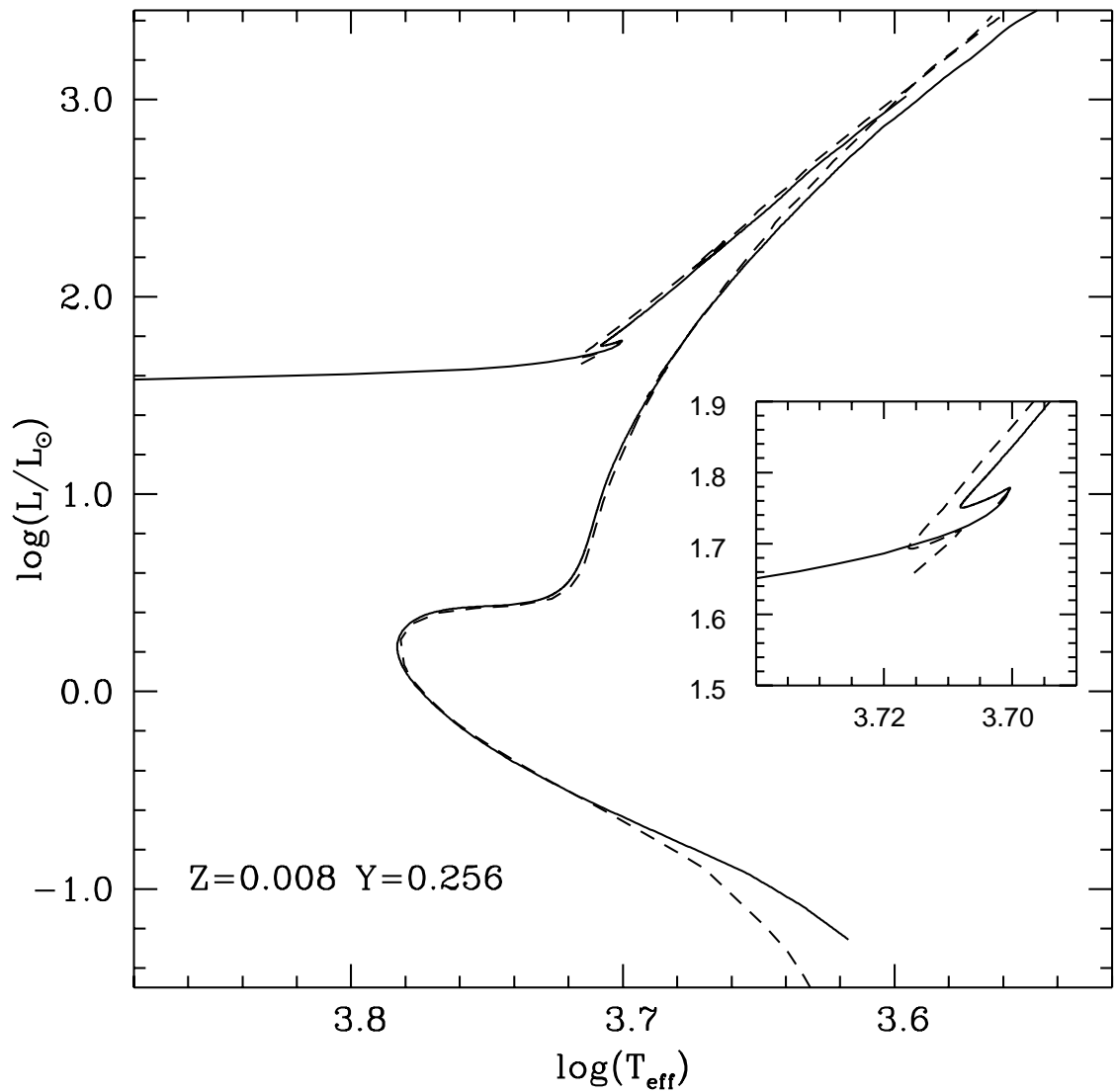}        
\caption{Comparison between our 12~Gyr old isochrone (solid line) with the
labeled chemical composition, and a 12~Gyr old, $Z=0.008$
$\alpha$-enhanced isochrone (dashed line) from Salasnich et
al~(2000). The inset shows an enlargement of the ZAHB region.
\label{Salas1}}        
\end{figure}

A similar comparison but with a 12~Gyr old $\alpha$-enhanced isochrones
by Salasnich et al.~(2000 - hereafter S00) is displayed in
Fig.~\ref{Salas1}. S00 models have also been computed using the same
metal mixture and opacities employed in our own calculations; at
$Z=0.008$ the initial He abundance of their models ($Y=0.250$) is close
enough to ours, to provide a meaningful comparison.
As for our models, we
display the ZAHB and the full 12~Gyr isochrone computed with the Reimers
parameter $\eta$=0.2. 

The effective temperatures agree well along MS and RGB for
$T_{eff}$ larger than about 5000~K. At lower temperatures the S00
isochrone is systematically hotter. The ZAHB luminosity of S00 models
is fainter by $\Delta\log(L/L_{\odot}) \sim$0.04.

\begin{figure}        
\plotone{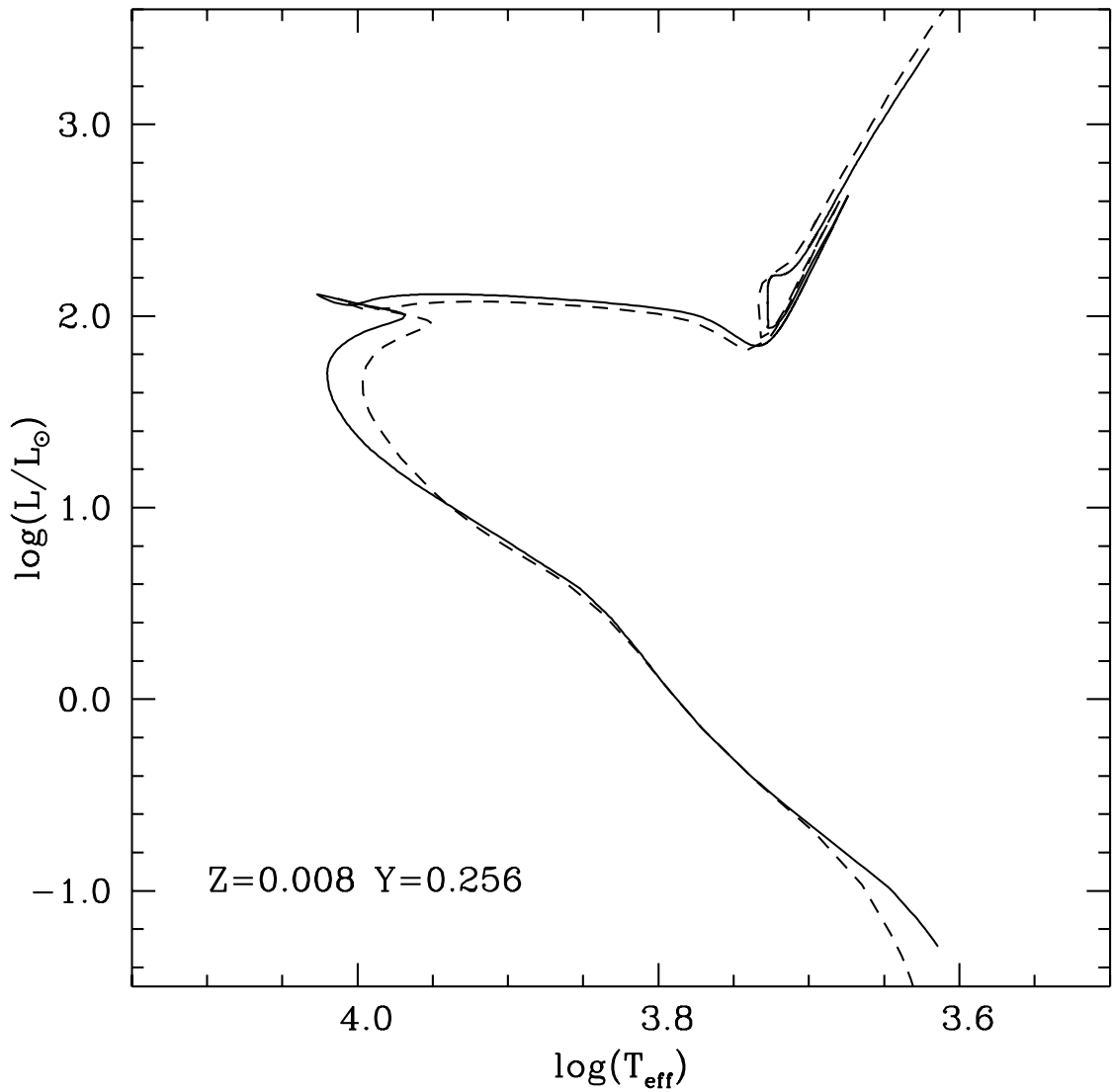}        
\caption{As in Fig.~\ref{Salas1}, but for an age of 500~Myr.
\label{Salas2}}        
\end{figure}

Given that S00 provide also $\alpha$-enhanced isochrones for low ages,
we show in Fig.~\ref{Salas2} a comparison of 500~Myr old
isochrones. Current and S00  isochrone include
overshooting from the convective core, albeit with different
prescriptions.
The TO region of our isochrone is brighter and hotter than S00
counterpart. This is consistent with the result of a similar
comparison between our 500~Myr scaled-solar isochrone of solar
metallicity and the corresponding one from Girardi et al.~(2000), that
should be computed with physical prescriptions very similar to S00. As
we remarked in Paper~I, the interplay between the different treatment of
the core overshooting and some different input physics should play a major role
in causing this difference. As for the scaled solar case our He
burning phase is brighter than S00 one, but the post-MS effective
temperatures of our isochrones are slightly cooler than S00,
whereas in the scaled solar case they were hotter.

\begin{figure}        
\plotone{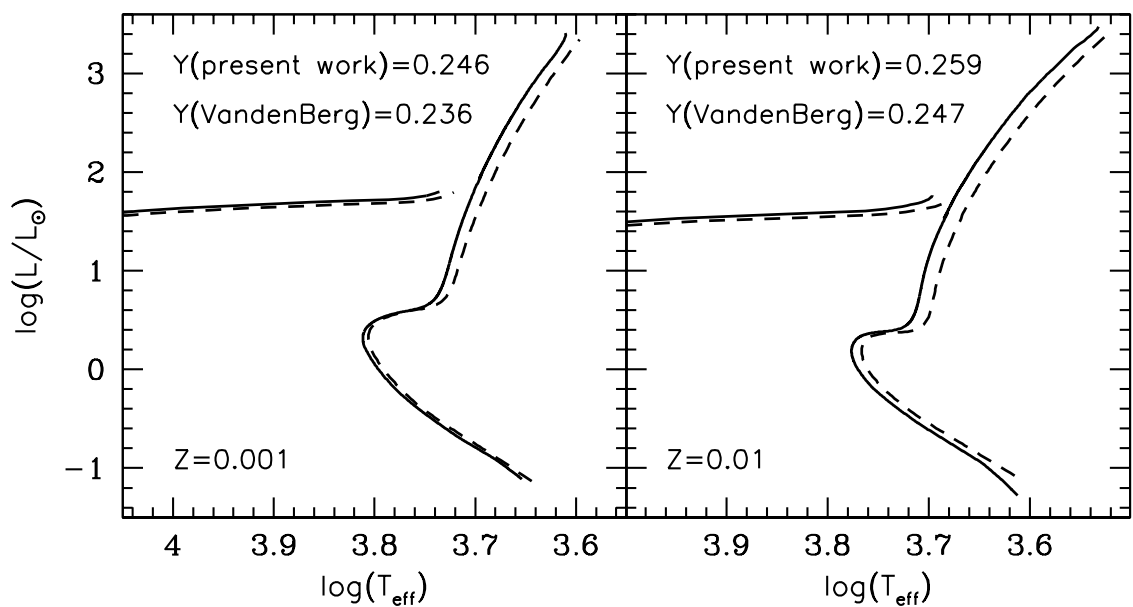}        
\caption{{\sl Left panel}: comparison between our 12~Gyr old isochrone (solid line) with the
labeled chemical composition, and a 12~Gyr old isochrone 
(dashed line) from Vandenberg et al.~(2000) with the same $Z$. {\sl
Right panel}: as the left panel but for $Z=0.01$. 
\label{dv}}        
\end{figure}

Figure~\ref{dv} shows the comparison with a 12~Gyr old isochrone by
VandenBerg et al.~(2000) for $Z$=0.001 and $Z$=0.01.  
Although the selected $Z$ is the same of our isochrones, 
there exists some difference at the level of 0.01 in the initial He
mass fraction, and the average $\alpha-$enhancement of our models is
equal to 0.4~dex, whereas the Vandenberg et al.~(2000) isochrones
displayed in this comparison have [$\alpha$/Fe]=0.3, all
$\alpha$-element being enhanced by the same amount. 

Bearing in mind these differences between the two data sets, 
one notices that at $Z=0.001$ the MS almost
completely overlap, whereas the TO of our isochrone is
slightly brighter, by $\Delta\log(L/L_\odot)\sim0.005$ and $\sim 80$~K
hotter. Along the RGB our isochrone is hotter by about 140~K, a 
difference larger than what we found in the case of the scaled-solar
models compared in Paper~I. This occurrence
cannot be attributed to a combination of different EOS and solar
mixing-length 
calibration, since we found
a very good agreement for scaled-solar RGBs. Perhaps, it is mainly an
effect due to the different
$\alpha-$element enhancement and mixture and, to a lesser extent, the
higher initial He content of our models. 
The brightness 
difference at $\log(T_{eff})=3.85$ along the ZAHB - taken as
representative of the mean effective temperature of the RR Lyrae instability
strip - is of the order of $\Delta\log(L/L_\odot)\sim0.04$.
 
The differences between the two sets of isochrones increase for
increasing $Z$. At $Z=0.01$ the TO of our isochrone is brighter by 
$\Delta\log(L/L_\odot)\sim0.009$ and $\sim140$~K hotter. 
Our RGB effective temperature is hotter by $\sim170$~K,
while the brightness difference at the RR Lyrae instability strip is equal to 
$\Delta\log(L/L_\odot)\sim0.05$. 

\begin{figure}        
\plotone{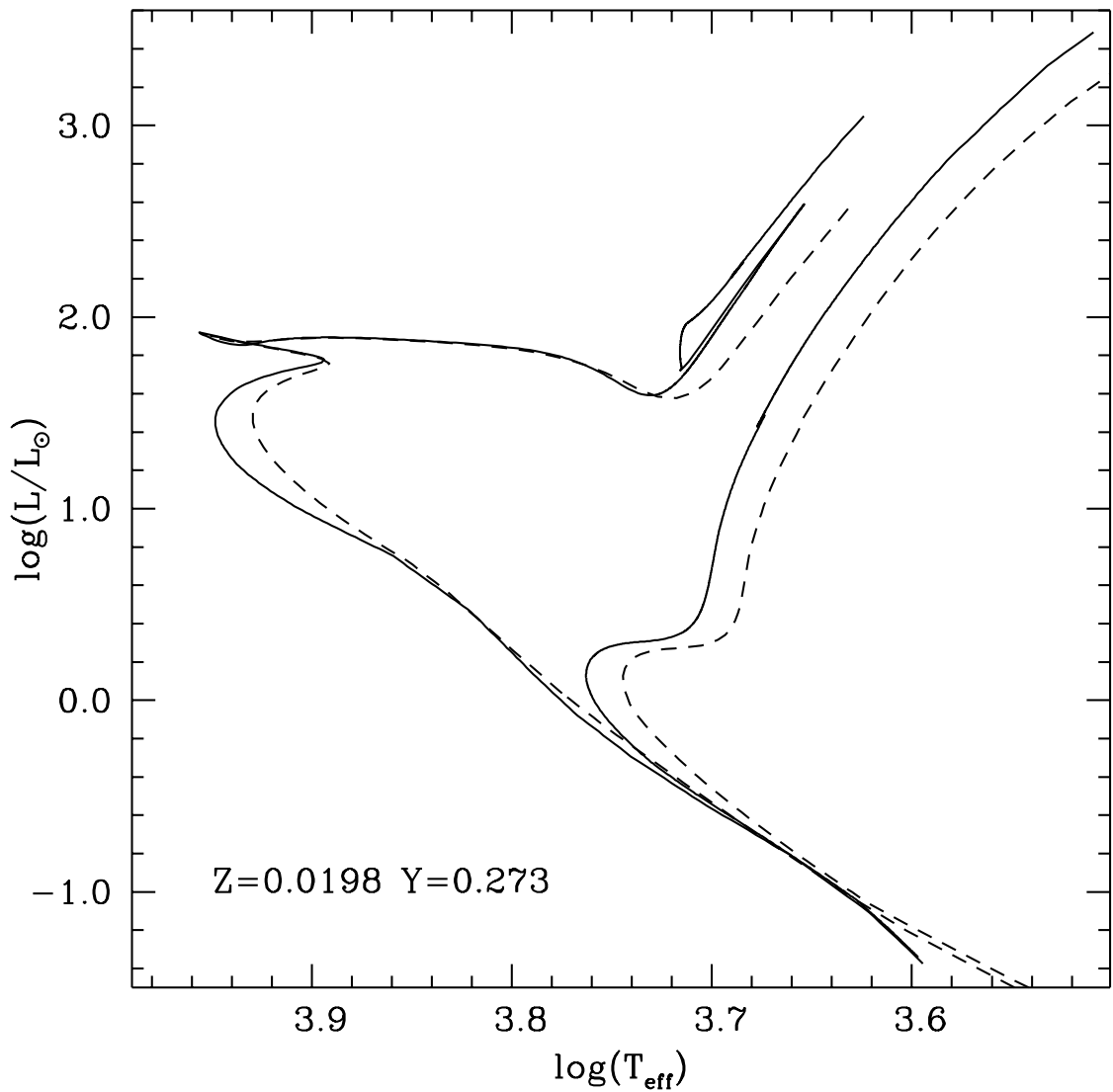}        
\caption{Comparison between our 13~Gyr and 600~Myr old isochrones for
the labeled composition, with the $Y$=0.270, $Z$=0.02, 
[$\alpha$/Fe]=0.3 isochrones of the same ages by Kim et al.~(2002).
\label{YY}}        
\end{figure}

Finally, we show in Fig.~\ref{YY} a comparison with the
$\alpha$-enhanced isochrones by Kim et al.~(2002), with
[$\alpha$/Fe]=0.3, $Z$=0.02 and $Y$=0.27. The $\alpha$-enhanced metal
distribution of Kim et al.~(2002) is the same as VandenBerg et
al.~(2000). We selected two ages, 
600~Myr and 13~Gyr, respectively.
Our 600~Myr old isochrone 
is computed with convective overshoot, given that Kim et al.~(2002)
models include overshoot from the convective core. The 
outcome of the comparison is very
similar to what we found in Paper~I in case of the Yi et al.~(2001)
scaled-solar models, which are homogeneous with the Kim et al.~(2002) library.
The MS appear in good agreement, apart from the lowest luminosities,
and the TO brightness of the 600~Myr isochrones are very similar,
although our TO is hotter. The RGBs of our isochrones are
systematically hotter, by about 200~K, similar to 
what we found for the scaled solar counterparts. The TO region of the 13~Gyr isochrone
from Kim et al.~(2002) is fainter and cooler, mainly due to the
effect of He-diffusion.


\section{Empirical tests}

\noindent
In this section we present a set of empirical tests, to  
show the level of agreement between our models and a number of 
photometric constraints. 

Our $\alpha$-enhanced isochrones have been
already compared to globular cluster (GC) CMDs in 
the ACS filters (Bedin
et al.~2005) and employed to determine the initial He content (from
the theoretical calibration of the $R$-parameter), 
distances (from the theoretical ZAHB models), 
reddening and relative ages of a large sample of
Galactic GCs (Salaris et al.~2004, Recio-Blanco et al.~2005 and De
Angeli et al.~2005). More in detail, the distances obtained from
fitting our ZAHB models to the observed HBs of a sample of Galactic
GCs (in the HST flight system) 
provides a distance scale in very good agreement with the empirical
MS-fitting distances determined by Carretta et al.~(2000), who employed a
sample of field subdwarfs with accurate Hipparcos parallaxes and
metallicity determinations. The relationship between $M_V$ of the ZAHB
and [Fe/H] from our models (see Table~3) is in agreement with the
similar relationship calibrated by Carretta et al.~(2000) from 
their empirical GC distance, within their quoted error bars. 
Moreover, the initial He content ($Y\sim
0.250$) obtained for the large sample of GCs analyzed by Salaris et
al.~(2004), is in line with predictions from primordial 
nucleosynthesis calculations, when 
using the value of the cosmological baryon density
estimated from WMAP data (Spergel et al.~2003).
In the following, we will make use of the extinction laws by Schlegel,
Finkbeiner \& Davis~(1998).

\subsection{Individual stars and globular clusters}

\begin{figure}        
\plotone{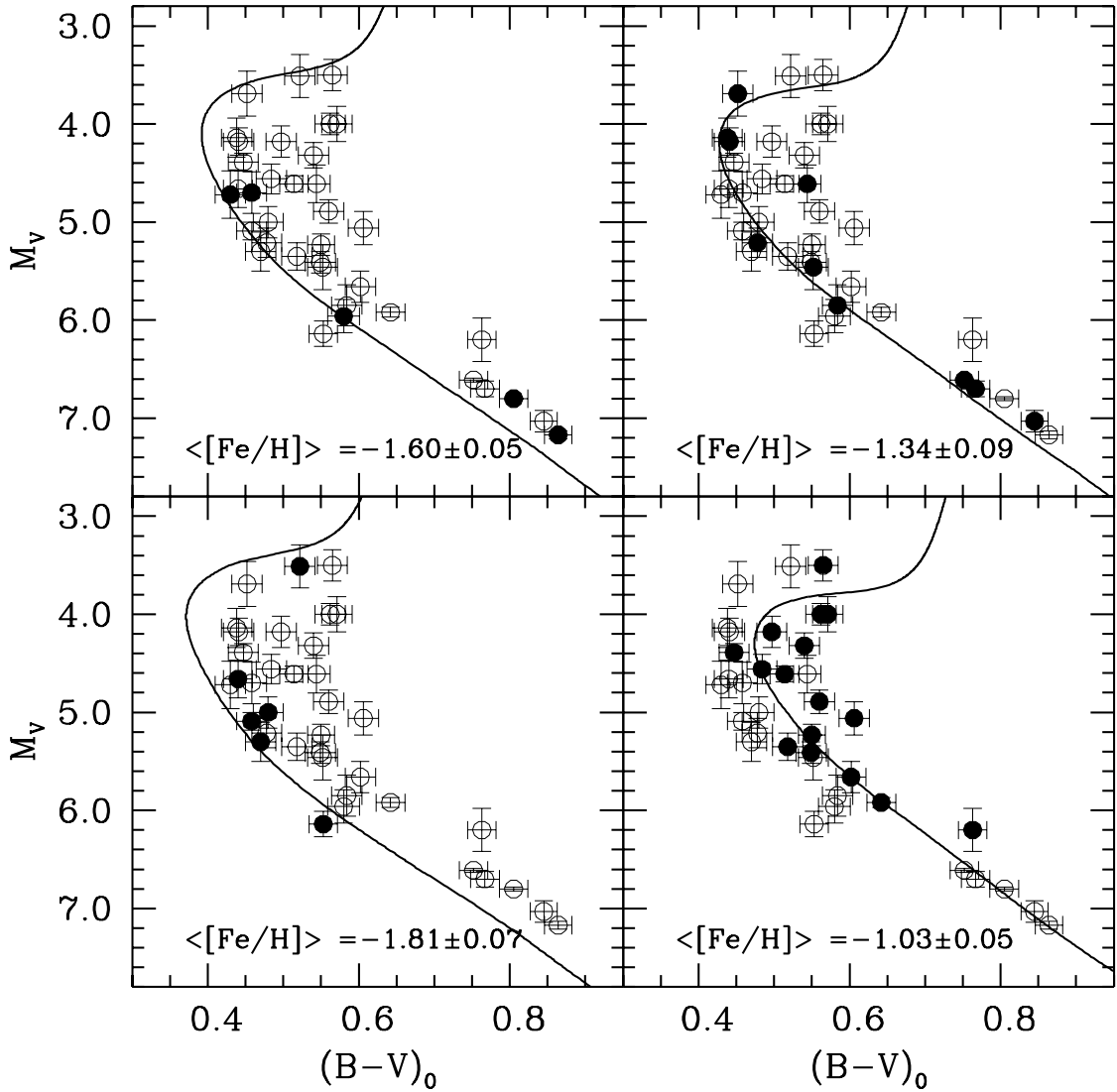}        
\caption{Comparison of the CMD of a sample of field subdwarfs with
Hipparcos parallaxes, metallicity and reddening determinations from
Carretta et al.~(2000), and our $\alpha$-enhanced isochrones. The full
sample is divided into four subsamples (displayed in each panel with
filled circles) with the labeled average [Fe/H]. Isochrones from our
database with age of 13~Gyr and [Fe/H] equal to $-$1.85, 
$-$1.62, $-$1.31 and $-$1.01 are displayed.
\label{subdw}}        
\end{figure}

Figure~\ref{subdw} shows the comparison of the CMD of the field subdwarf sample by
Carretta et al.~(2000) with isochrones of the appropriate
metallicity. These objects have accurate Hipparcos parallaxes, 
spectroscopic metallicities, individual reddening determinations, 
and have been grouped into four subsamples of different mean [Fe/H]. 
We have compared the CMD of these objects with our
13~Gyr old isochrones, whose [Fe/H] is within at most 0.04~dex
from the mean values displayed in the figure. The precise choice of the
isochrone age does not play any role when $M_V\ge5.5$~mag.

Theory and observations appear in satisfactory agreement, particularly
in the magnitude range unaffected by age. 
At $M_V$=6 the theoretical isochrones
trace very well the change of subdwarf colors with changing metallicity. At
[Fe/H]$\sim -$1.60 and $\sim -$1.0 the faintest subdwarfs appear to be
systematically redder than the models, but at [Fe/H]$\sim -$1.3 they
agree within the quoted 1$\sigma$ error bars.

\begin{figure}        
\plotone{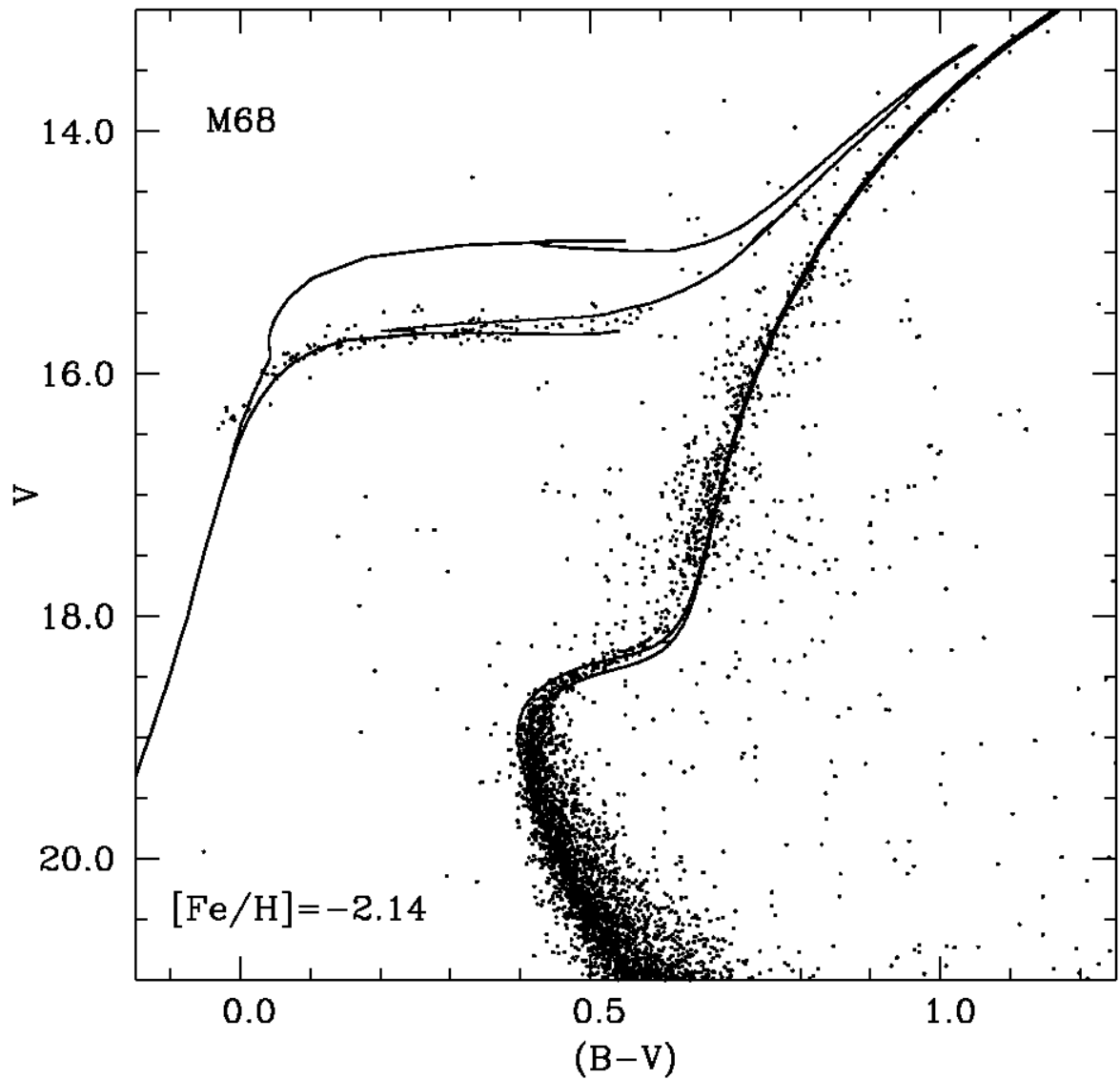}        
\caption{$BV$ diagram of M68 (data from Walker~1994) compared to [Fe/H]=$-$2.14, 11 and 12~Gyr old
isochrones. See text for details.
\label{M68}}        
\end{figure}

\begin{figure}        
\plotone{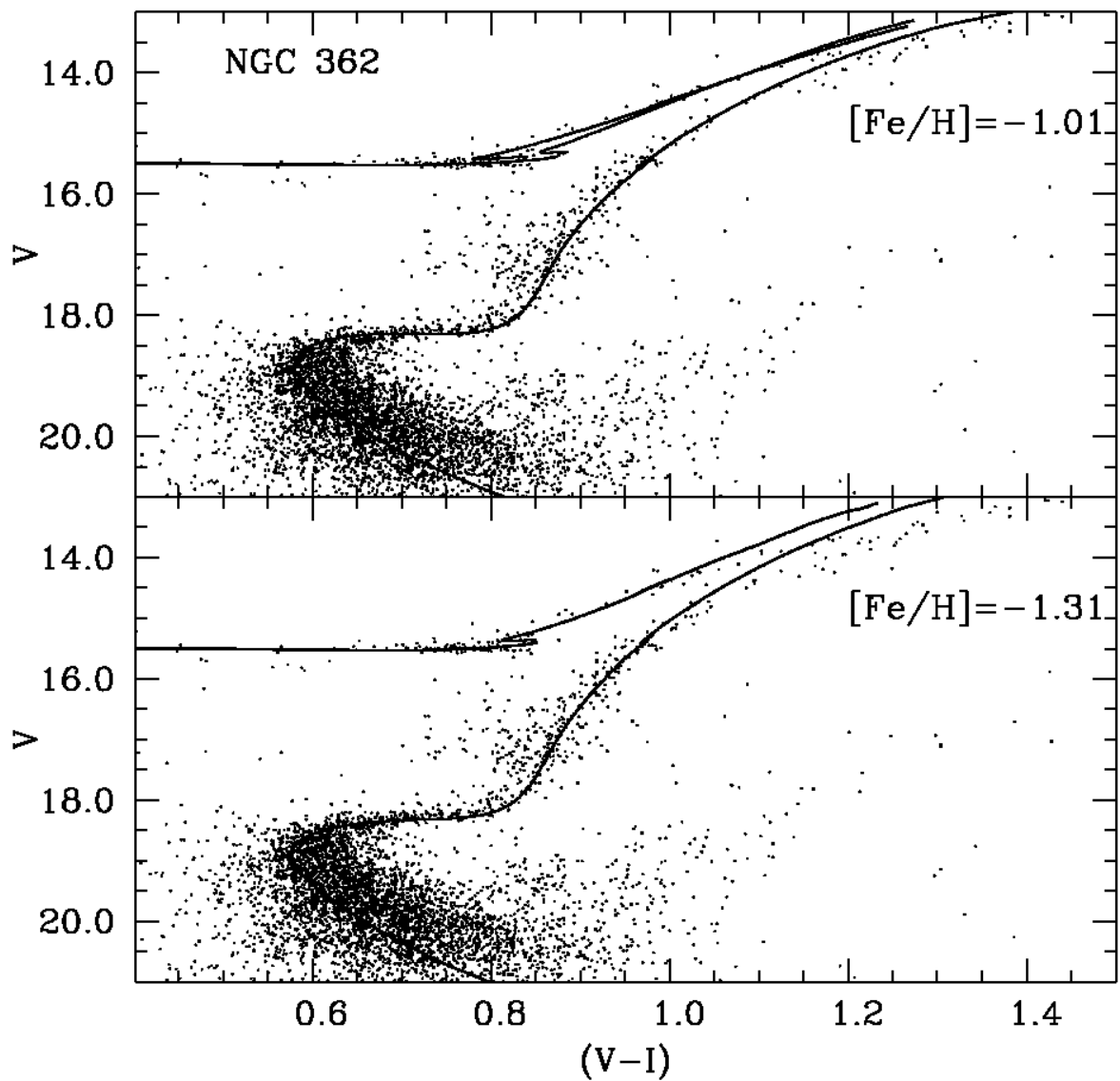}        
\caption{$VI$ diagram of NGC~362 (data from Bellazzini et al.~2001) compared to isochrones with,
respectively, [Fe/H]=$-$1.31, t=10~Gyr and [Fe/H]=$-$1.01,
t=9~Gyr. See text for details. 
\label{N362}}        
\end{figure}

The figures~\ref{M68} and~\ref{N362} show the comparison of our isochrones with the
CMD of two Galactic GCs, namely $BV$ data for M68 (data from
Walker~1994) and $VI$ data for NGC~362
(data from Bellazzini et al.~2001). A detailed analysis of the fitting
to globular cluster CMDs and the study of their age distribution is
outside the scope of the paper. As mentioned before, here
we wish only to show the level of agreement between our set of models
and selected CMDs of Galactic GCs of different [Fe/H] values.

For M68 we display in Fig.~\ref{M68} our 11 and 12~Gyr 
old isochrones with [Fe/H]=$-$2.14, corrected
for a reddening $E(B-V)=$0.07 and distance modulus
$(m-M)_V$=15.25. The chosen metallicity is somewhat intermediate among the
values [Fe/H]=$-1.99$ on the Carretta \& Gratton~(1997
-- CG97) scale, [Fe/H]=$-2.09$ on the Zinn \& West~(1984 -- ZW84) scale and
[Fe/H]=$-$2.40 on the Kraft \& Ivans~(2003 -- KI03) scale.
The adopted reddening (necessary to match the location of the observed MS) 
is in good agreement with the value 
$E(B-V)=$0.07$\pm$0.01 estimated by Walker~(1994).
The distance is obtained from the fitting of the theoretical ZAHB to
the observed counterpart, and agrees with the value we determined in
the HST system (Recio-Blanco et al.~2004).
Together with the predicted ZAHB we also display the HB and AGB
portion of the 12~Gyr isochrone, for values of the mass loss 
parameter $\eta$=0.2 and 0.4, respectively.

Figure~\ref{N362} displays the comparison with NGC362. 
NGC~362 has a metallicity [Fe/H]=$-$1.27 on both ZW84 
and KI03 scales, and [Fe/H]=$-$1.15 on the CG97 scale; we 
compare the observed CMD to our isochrones with [Fe/H]=$-$1.31 (t=10~Gyr
$E(B-V)=$0.02) and an isochrone with [Fe/H]=$-$1.01 (t=9~Gyr, $E(B-V)=$0.0). 
Together with the full theoretical ZAHB we display the HB and AGB
portion of the isochrones computed with $\eta=0.2$ for 
[Fe/H]=$-$1.31, and both  $\eta$=0.2 and 0.4 for [Fe/H]=$-$1.01.
The distance moduli adopted from the fitting of the observed 
ZAHB are $(m-M)_V=14.94$ for 
[Fe/H]=$-$1.31 and $(m-M)_V$=14.86 for [Fe/H]=$-$1.01, consistent 
with Recio-Blanco et al.~(2004 -- obtained from our models transformed
into the HST flight system) and very close to the MS-fitting distance by Carretta
et al.~(2000 -- $(m-M)_V=14.98 \pm 0.05$).

\begin{figure}        
\plotone{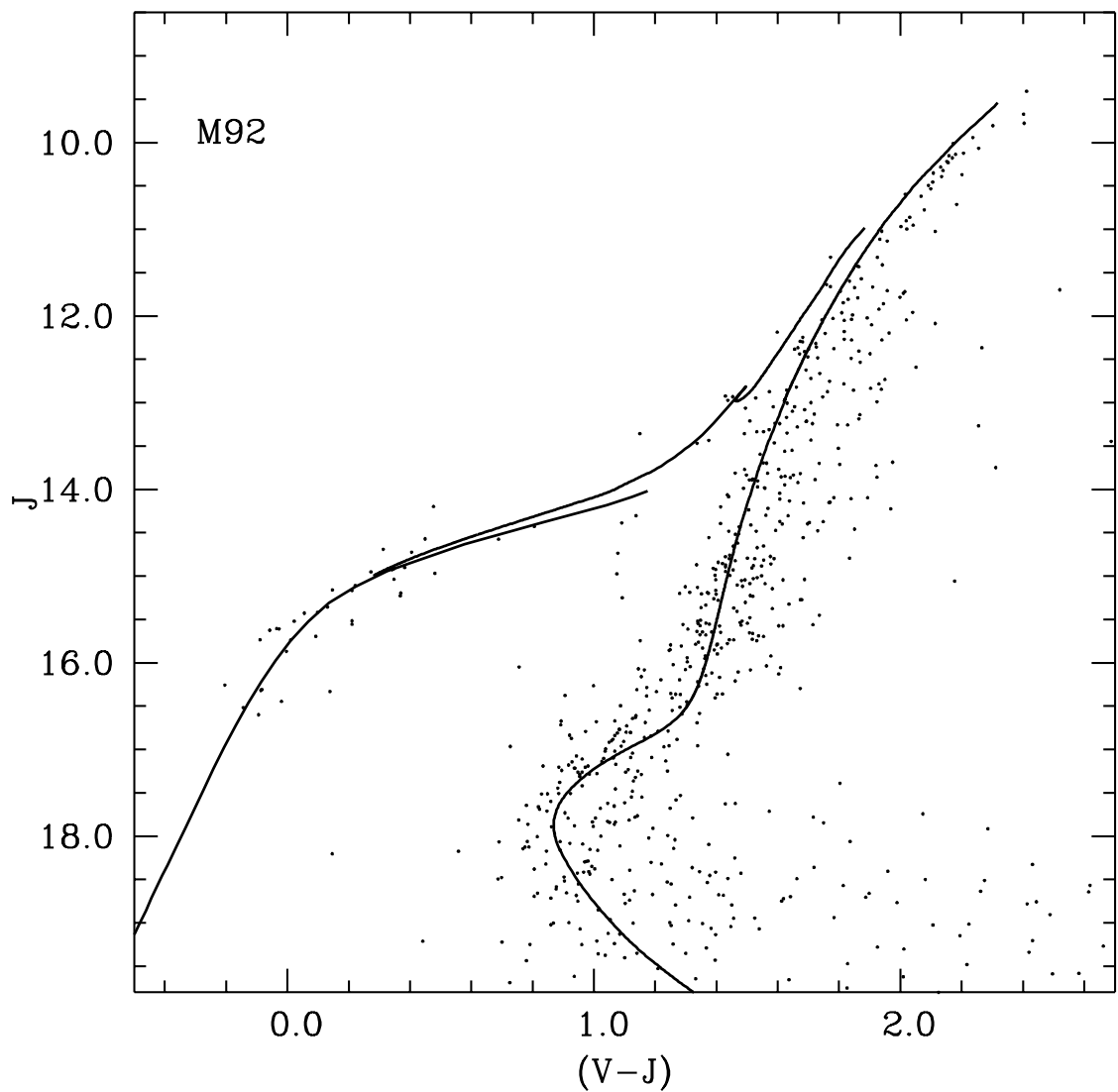}        
\caption{$VJ$ diagram of M92 (data from Del Principe et al.~2005) compared to an isochrone with
[Fe/H]=$-$2.14, t=13~Gyr. See text for details.
\label{M92}}        
\end{figure}

We have also performed some comparisons with near infrared data of
Galactic globular clusters. Figure~\ref{M92} shows our 13~Gyr old isochrone 
with [Fe/H]=$-$2.14, including the full ZAHB plus the HB and AGB
evolution computed with $\eta$=0.2, together with a $VJ$ CMD of
M92 (Del Principe et al.~2005). We recall that 
M92 has [Fe/H]=$-$2.16 on the CG97 scale
and [Fe/H]=$-$2.24 on the ZW84 scale.
We have shifted the models to account
for the well established reddening $E(B-V)=$0.02 (Schlegel et al.~1998) 
and a distance modulus $(m-M)_J=14.75$,
that corresponds to $(m-M)_0=14.72$. This value is within $\sim 
1\sigma$ of the empirical MS-fitting distance 
obtained by Carretta et al.~(2000), e.g., $(m-M)_0=14.64\pm 0.07$.
The dispersion of the observational CMD does not allow us to firmly constrain the 
derived distance and age; nevertheless the
displayed isochrone properly fits the various branches of the observed diagram.

\begin{figure}        
\plotone{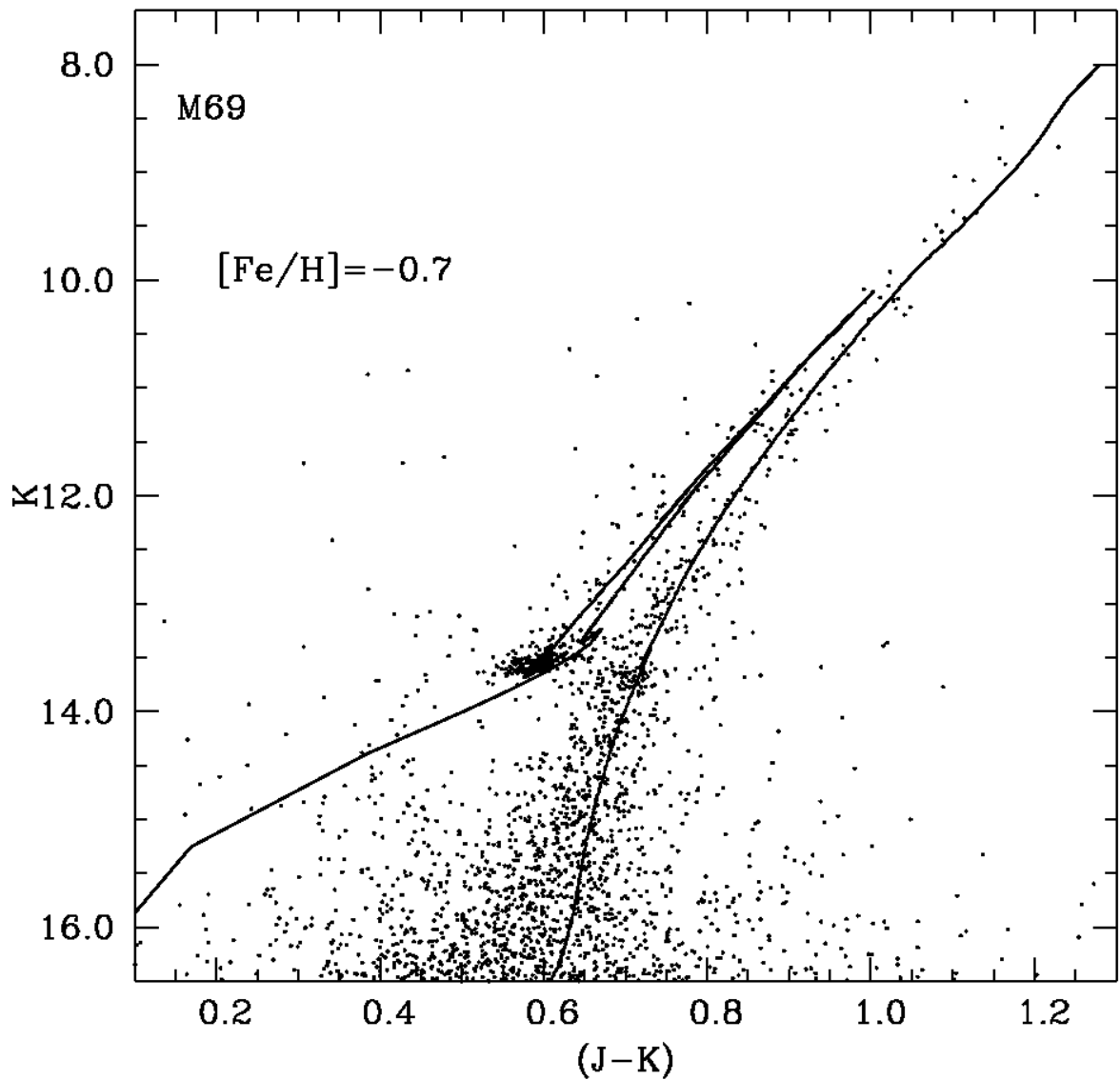}        
\caption{$JK$ diagram of M69 (Valenti et al.~2005) compared to a [Fe/H]=$-$0.7, t=10~Gyr
isochrone. See text for details.
\label{M69}}        
\end{figure}

A $JK$ CMD of the HB and bright RGB of the [Fe/H]$\sim -0.7$ cluster M69 
(Valenti, Origlia \& Ferraro~2005) is shown in Figure~\ref{M69},
compared to our 10~Gyr, [Fe/H]=$-0.7$ isochrone (including ZAHB
plus the HB and AGB evolution for both $\eta$=0.2 and 0.4). 
The observational
data were on the 2MASS system and have been transformed to the $JK$ 
system consistent with our color transformations using the empirical
relationships by Carpenter~(2001). 
We have adopted a distance modulus $(m-M)_0=14.76$ from fitting the
theoretical ZAHB to the observed HB, and a reddening $E(B-V)$=0.15; 
both values are very close to $(m-M)_0=14.78$ and 
$E(B-V)$=0.16 listed in the Harris~(1996) catalog.
The displayed isochrone traces very well the position of the
RGB up to the brightest detected stars.

\begin{figure}        
\plotone{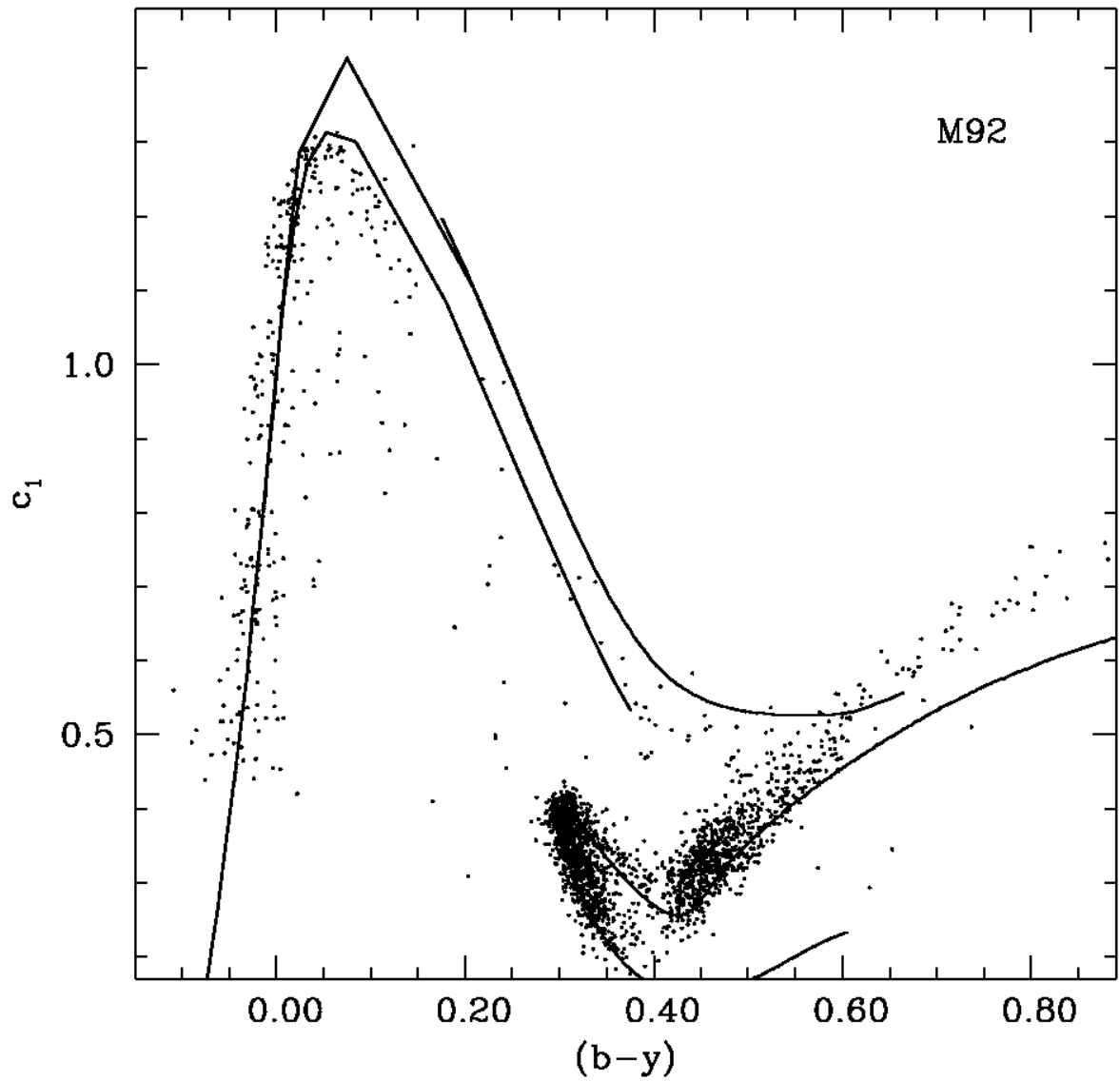}        
\caption{The $[c_1, b-y]$ diagram of M92 (data from Grundahl et al.~2000), compared to
a theoretical isochrone with t=13~Gyr,
[Fe/H]=$-$2.14. See text for details.
\label{M92Strom}}        
\end{figure}

\begin{figure}        
\plotone{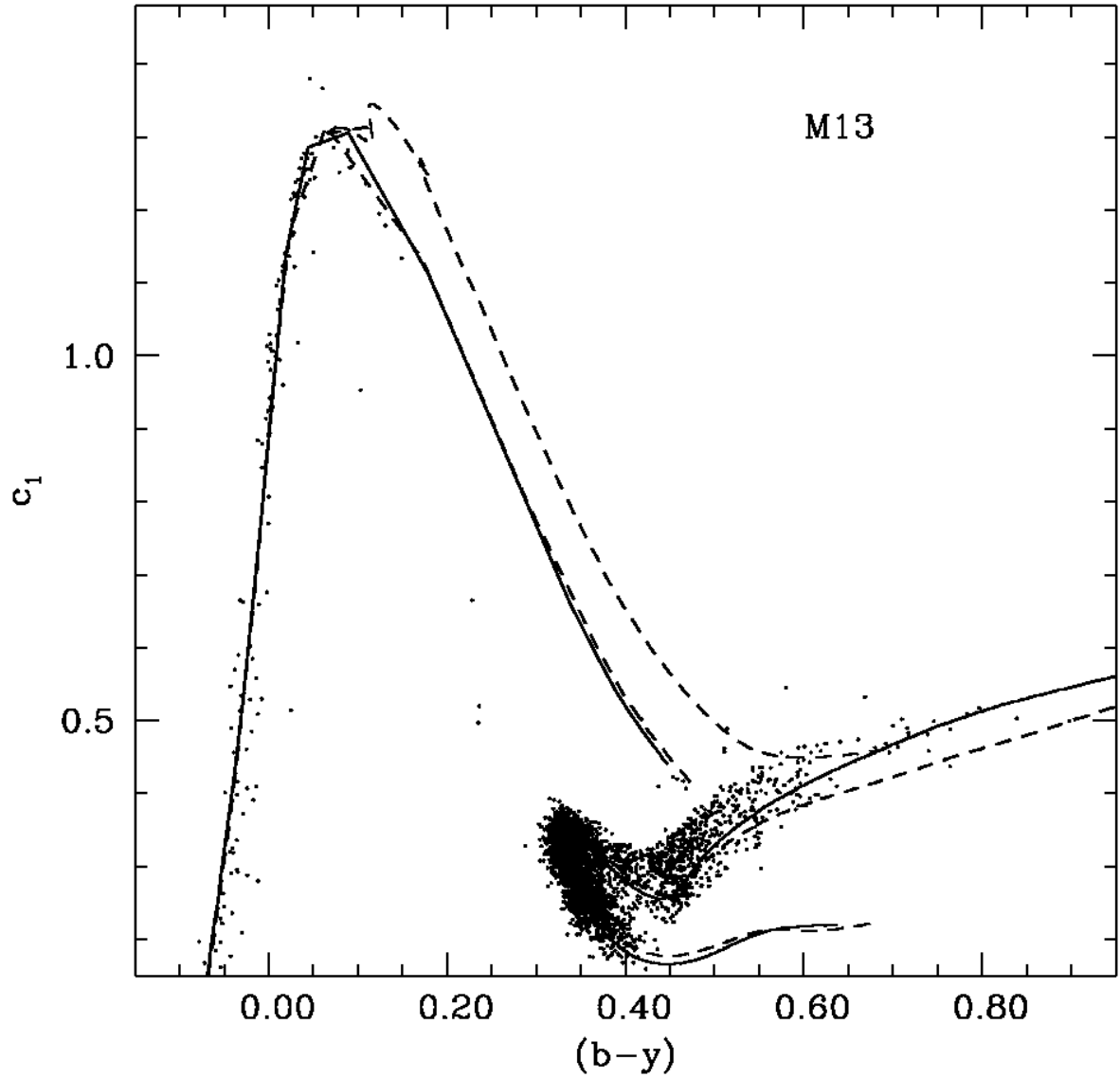}        
\caption{The $[c_1, b-y]$ diagram of M13 (Grundahl et al.~1998), compared to
two theoretical isochrones with, respectively, t=13~Gyr,
[Fe/H]=$-$1.62 (solid line), and t=11~Gyr, [Fe/H]=$-$1.31 (dashed line). See text for details.
\label{M13Strom}}        
\end{figure}

Figures~\ref{M92Strom} and \ref{M13Strom} show the comparison of
our isochrones with the Str\"omgren $[c_1,b-y]$ diagrams of M92 and
M13 (Grundahl et al.~1998, 2000). In case of M92 we show a 
13~Gyr old [Fe/H]=$-$2.14 isochrone computed with $\eta=0.4$, together with the
full ZAHB. The only correction applied to the models is for the
reddening $E(B-V)$=0.02, where we have used the relationships
$E(b-y)=E(B-V)/1.35$, and $c_1=c_0+0.2E(b-y)$. 
Notice how the ZAHB sequence runs through the center of the observed
HB distribution. This is exactly what one expects from theory, 
given that the evolution off the ZAHB along 
the blue part of the HB on this $[c_1,b-y]$ diagram overlaps with the
ZAHB itself. When the ZAHB sequence turns toward the red,  
theoretical models show a value of $c_1$ slightly larger than
observed. The isochrones overlap to various degrees all other 
main branches of the observed diagram, including the RGB. 
One has also to notice the spread along the observed RGB, due
to the large scatter in $c_1$ caused by star-to-star differences in
the abundance of nitrogen (see, e.g. Grundahl et al.~1998).

In case of M13 (see Fig.~\ref{M13Strom}) we have displayed two 
isochrones and ZAHBs with, respectively,
t=13~Gyr, [Fe/H]=$-$1.62, and t=11~Gyr, [Fe/H]=$-$1.31. This accounts
for the uncertainty in M13 iron content, which is [Fe/H]=$-1.39$ according
to CG97, [Fe/H]$\sim -$1.5 according to KI03 and [Fe/H]=$-1.65$ according to ZW84.
Only for the [Fe/H]=$-$1.62 isochrone we have displayed the full HB
evolution obtained using the Reimers parameter $\eta$=0.4 along
the RGB. 
The models have been shifted for a reddening $E(B-V)$=0.03, very close
to the value $E(B-V)$=0.02 given by Schlegel et al.~(1998).
The level of agreement between theory and observations is similar to
the case of M92. Notice again the wide RGB, as for M92 (see Fig.~\ref{M92Strom}).

\subsection{Integrated photometric properties}

We show here a few comparisons of integrated properties extracted from
our isochrone sets, e.g. Surface Brightness Fluctuations (SBFs) in
$V$ and $I$, plus $(B-V)$ and $(V-I)$ integrated colors of a sample
of Galactic GCs. They can be considered as a global test of the
accuracy of the bolometric corrections, colors and evolutionary 
timescales along the relevant phases.
We selected GCs whose SBFs in $V$ and $I$ have been
determined by Ajhar \& Tonry~(1994), and whose distances have been
estimated by Recio-Blanco et al.~(2005) using our $\alpha$-enhanced
ZAHB models. 
As for the integrated colors, we considered the clusters with both
$(V-I)$ and $(B-V)$ integrated photometry plus reddenings and [Fe/H] 
values available in the latest version of the Harris~(1996)
catalog, which are an average of various sources.

It is important to remark that the $V$ and $I$ SBFs, and the
integrated $(B-V)$ and $(V-I)$ colors are very weakly affected by the 
full thermal pulse phase and by MS masses below 0.5$M_{\odot}$
(presently not included in our database -- see, e.g., Brocato et
al.~2000, Cantiello et al.~2003).

The theoretical fluctuation magnitudes have 
been as usual computed from the fluctuation luminosity $\bar{L}$ in the 
selected photometric band, given by 

$$\bar{L}=\frac{\sum n_i L_i^2}{\sum n_i L_i}$$

where $n_i$ is the number of stars of type $i$ and luminosity $L_i$.
These sums have been performed along the appropriate isochrone, 
populated according to a Salpeter initial mass function (the
integrated colors have also been computed using the same initial mass function). 

\begin{figure}        
\plotone{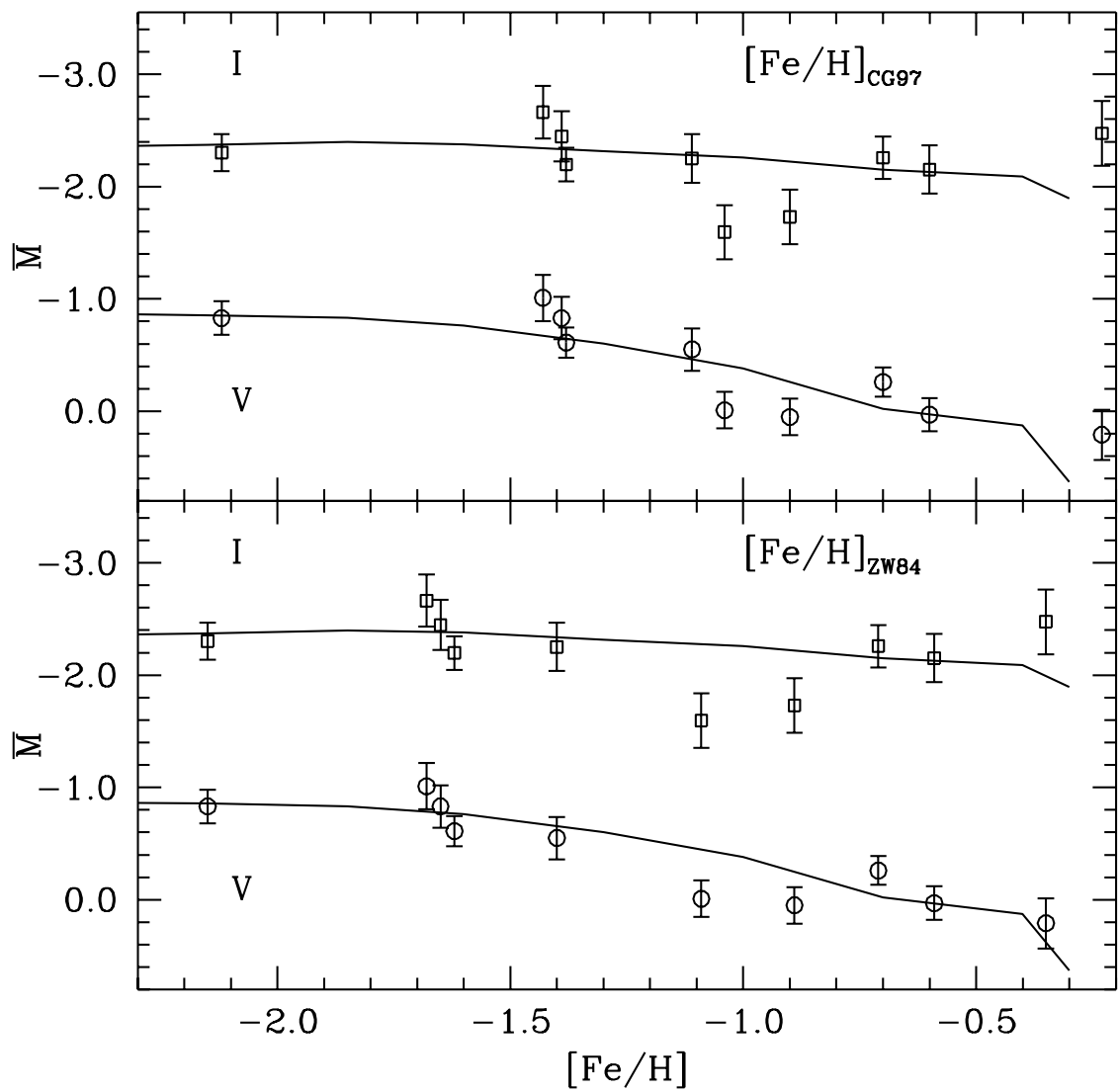}        
\caption{$V$ and $I$ fluctuation magnitudes for a sample of Galactic
GCs (Ajhar \& Tonry~1994) compared to the theoretical predictions. See
text for details. 
\label{SBF}}        
\end{figure}

Figure~\ref{SBF} shows the comparison between the observed GC $\bar{I}$ and $\bar{V}$ 
fluctuation magnitudes, and
the theoretical counterpart obtained from our 10~Gyr $\eta=0.2$
isochrones. In order to show the effect of the uncertain [Fe/H] scale we
have considered both the ZW84 and CG97 values. 
The 1$\sigma$ error bars of the observational points have been obtained by
adding in quadrature the 
1$\sigma$ errors given by Ajhar \& Tonry~(1994) and the errors on the
individual distance moduli given by Recio-Blanco et al.~(2005).
The theoretical fluctuation magnitudes trace very well the
observations; the effect of a larger age (e.g. 13 instead of 10~Gyr)
and a bluer HB ($\eta$=0.4 instead of $\eta$=0.2) are negligible on
this plot. 
As noticed by our referee, 
two clusters, namely NGC~6652 and NGC~6723 (with [Fe/H]$\sim -$1.0)   
are systematically discrepant in both the $\bar{V}$ and $\bar{I}$ fluctuation
magnitudes, by about 0.2-0.4~mag.
It is difficult to point out a precise reason for this
discrepancy --  which is at the level of at most 
$\sim 3 \sigma$ -- although we believe it is possible to exclude an 
underestimate of the distances to these two clusters by such a large amount. 
Here we just notice that a similar discrepancy in
$\bar{V}$ and $\bar{I}$ can be
found in the analysis by Cantiello et al.~(2003 -- their Fig.~12), 
who use their own SBF models, [Fe/H] and distances from the Harris~(1996) catalog. 

\begin{figure}        
\plotone{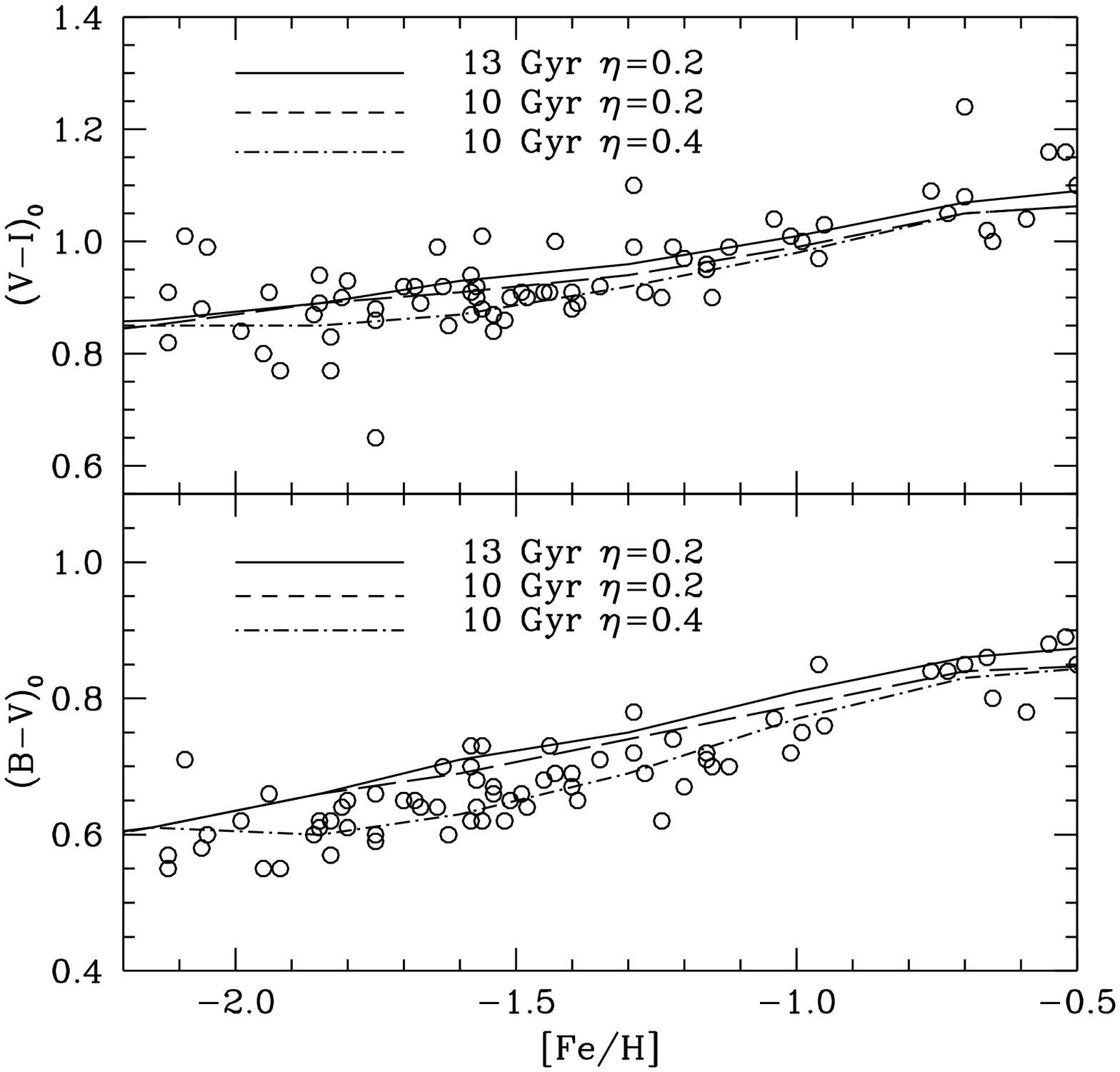}        
\caption{Integrated colors of Galactic GCs from Harris~(1996)
compared to the theoretical predictions. See text for details.
\label{intcol}}        
\end{figure}

A comparison with the observed $(V-I)$ and $(B-V)$ integrated colors 
is displayed in Fig.~\ref{intcol}. The
observed colors have been dereddened using the $E(B-V)$ values given
in the Harris' catalog. Given the heterogeneous sources for the
observational data one can only discuss the general agreement with
theory.
Not surprisingly, the integrated $(V-I)$ is
very weakly sensitive to age and also to the assumed value of $\eta$,
being mainly affected by [Fe/H]. Our models trace very well the
observed trend of $(V-I)$ with [Fe/H]. 
The $(B-V)$ color is more sensitive to both the age and HB
morphology; even in this case the models reproduce well the
observations. 

\section{Summary}
 
This paper presents an extension of the scaled-solar stellar model and
isochrone database of Paper~I, to include an $\alpha$-enhanced
([$\alpha$/Fe]=+0.40) metal distribution.

The models have been computed by using the same sources for the input
physics and the same stellar evolution code of Paper~I. They are based
on the most updated stellar physics available, and are fully
consistent with our scaled-solar database. This allows one to
disentangle unambiguously the effect of different heavy element
distributions when analizing the properties of stellar populations. 

The database covers a wide metallicity range suitable for both metal-poor stars,
such as extreme Population~II objects, and metal-rich stars such as
the ones hosted by the
Galactic bulge and elliptical galaxies. The adopted initial He content
for the most metal-poor stellar models is 
consistent with recent predictions based on CMB analysis and the {\sl
R}-parameter investigations on a large sample
of Galactic GCs, and the adopted He-enrichment ratio accounts for the
initial solar He abundance based on our calibration of the Solar Standard Model.

Current database allows the computations of isochrone sets covering 
a wide range of stellar ages. We are already working to extend our computations 
to the end of the thermal pulse phase along the AGB.

For each chemical composition in our grid, we computed an extended set of HB models 
corresponding to an RGB progenitor whose age at the He flash is of the
order of 
$\sim13$~Gyr, i.e., suitable for
investigating the properties of HB stars in old stellar systems. So
far, this 
is the most complete
and homogeneous set of HB models for an $\alpha-$enhanced metal 
distribution, available in literature.

As in Paper~I, we devote great care to bolometric corrections and
color-$T_{eff}$ transformations. We have been able
to use CT based on a metal distribution consistent with 
that adopted for the evolutionary computations. This opportunity allowed us to extend
the investigation performed by Cassisi et al.~(2004) 
to the high metallicity regime and more narrow photometric bands such as the
Str{\"o}mgren filters. Our analysis makes more evident the need to
employ appropriate $\alpha$-enhanced CT, especially at high metallicities.

The evolutionary tracks and isochrones are normalized 
in such a way that they can be easily and directly included in any population
synthesis tool by requiring a minimum amount of work. 

We have discussed some relevant comparisons between our theoretical 
framework and empirical data  
for both unevolved field MS stars with accurate distance and metallicity estimates, and  
Galactic GCs with accurate photometry in various photometric systems. 
As a whole we notice an agreement between theory and observations.

The whole library is made available to the scientific community in an 
easy and direct way through a purpose built WEB site, the same that 
makes public our scaled-solar database. 
This WEB site is continuously updated to include additional
theoretical models, fully consistent with those presented here and in
Paper~I. In collaboration with D. Cordier,
a big effort is being made to prepare some WEB interactive interfaces that 
allow users to compute their desired isochrones,
evolutionary tracks, luminosity functions and synthetic CMDs.

\acknowledgments
We wish to warmly thank D. Cordier and M. Castellani for the relevant help
provided when preparing the BASTI WEB pages. We thank also M. Quintini for his
painstaking help in updating the BASTI database in the WEB server at the Observatory
of Teramo. It is a pleasure to thank G. Bono for a detailed reading of
an early draft of this paper.
This research has made use of NASA's Astrophysics Data System Abstract
Service and the SIMBAD database operated at CDS, Strasbourg, France.



\clearpage
\begin{deluxetable}{ccc}
\tablewidth{0pt}
\tablecaption{The adopted $\alpha$-enhanced heavy element mixture}
\tablehead{
\colhead{element} &
\colhead{Number fraction}&
\colhead{Mass fraction} }   
\startdata
  C     &  0.108211    &    0.076451 \nl
  N     &  0.028462    &    0.023450 \nl
  O     &  0.714945    &    0.672836 \nl
  Ne    &  0.071502    &    0.084869 \nl
  Na    &  0.000652    &    0.000882 \nl
  Mg    &  0.029125    &    0.041639 \nl
  Al    &  0.000900    &    0.001428 \nl
  Si    &  0.021591    &    0.035669 \nl
  P     &  0.000086    &    0.000157 \nl
  S     &  0.010575    &    0.019942 \nl
  Cl    &  0.000096    &    0.000201 \nl
  Ar    &  0.001010    &    0.002373 \nl
  K     &  0.000040    &    0.000092 \nl
  Ca    &  0.002210    &    0.005209 \nl
  Ti    &  0.000137    &    0.000387 \nl
  Cr    &  0.000145    &    0.000443 \nl
  Mn    &  0.000075    &    0.000242 \nl
  Fe    &  0.009642    &    0.031675 \nl
  Ni    &  0.000595    &    0.002056 \nl

\enddata 
\label{mixture} 
\end{deluxetable}

\begin{deluxetable}{cccr}      
\tablewidth{0pt}      
\tablecaption{Initial chemical compositions of our model grid.}      
\tablehead{      
\colhead{$Z$}&       
\colhead{$Y$}&
\colhead{[Fe/H]}&       
\colhead{[M/H]}}      
  \startdata   
      0.0001 &       0.245 & $-$2.62  & $-$2.27   \\
      0.0003 &       0.245 & $-$2.14  & $-$1.79   \\
      0.0006 &       0.246 & $-$1.84  & $-$1.49   \\
      0.0010 &       0.246 & $-$1.62  & $-$1.27   \\
      0.0020 &       0.248 & $-$1.31  & $-$0.96   \\        
      0.0040 &       0.251 & $-$1.01  & $-$0.66   \\
      0.0080 &       0.256 & $-$0.70  & $-$0.35   \\
      0.0100 &       0.259 & $-$0.60  & $-$0.25   \\
      0.0198 &       0.273 & $-$0.29  &    0.06   \\
      0.0300 &       0.288 & $-$0.09  &    0.26   \\
      0.0400 &       0.303 &    0.05  &    0.40   \\
\enddata      
\label{chemcomp}
\end{deluxetable}      

\begin{deluxetable}{cccccccc}      
\tablewidth{0pt}      
\tablecaption{Selected properties of ZAHB stellar models}      
\tablehead{      
\colhead{$Z$}& 
\colhead{$[Fe/H]$}&      
\colhead{$M_{pr}/M_\odot$\tablenotemark{a}}&
\colhead{$He_{sur} $\tablenotemark{b}}&       
\colhead{$M_{cHe}/M_\odot$\tablenotemark{c}}&       
\colhead{$M_{3.85}/M_\odot$\tablenotemark{d}}&       
\colhead{$log(L_{3.85}/L_\odot)$\tablenotemark{e}}&       
\colhead{$M_{V,3.85}$\tablenotemark{f}}
}      
\startdata   
      0.0001 & $-$2.62 & 0.80 & 0.253 & 0.5028 & 0.802 & 1.7573  &  0.375   \\
      0.0003 & $-$2.14 & 0.80 & 0.255 & 0.4969 & 0.710 & 1.7142  &  0.479   \\
      0.0006 & $-$1.84 & 0.80 & 0.258 & 0.4934 & 0.668 & 1.6955  &  0.518   \\
      0.0010 & $-$1.62 & 0.80 & 0.259 & 0.4905 & 0.642 & 1.6764  &  0.559   \\
      0.0020 & $-$1.31 & 0.80 & 0.262 & 0.4877 & 0.613 & 1.6510  &  0.615   \\
      0.0040 & $-$1.01 & 0.80 & 0.266 & 0.4845 & 0.589 & 1.6161  &  0.691   \\
      0.0080 & $-$0.70 & 0.90 & 0.276 & 0.4795 & 0.567 & 1.5705  &  0.812   \\
      0.0100 & $-$0.60 & 0.90 & 0.279 & 0.4781 & 0.561 & 1.5545  &  0.825   \\
      0.0198 & $-$0.29 & 1.00 & 0.296 & 0.4713 & 0.542 & 1.5025  &  0.948   \\
      0.0300 & $-$0.09 & 1.00 & 0.310 & 0.4656 & 0.529 & 1.4753  &  0.998   \\
      0.0400 &    0.05 & 1.00 & 0.324 & 0.4603 & 0.519 & 1.4657  &  1.015   \\
\enddata 
\tiny{
\tablenotetext{a}{Initial total mass of the RGB progenitor (in solar units).}
\tablenotetext{b}{Envelope He abundance at the He flash at the tip of the RGB.}
\tablenotetext{c}{He core mass at the core He-burning ignition (in solar units).}
\tablenotetext{d}{Mass of the stellar model whose ZAHB location is at $\log(T_{eff})=3.85$ (in solar units).}
\tablenotetext{e}{Logarithm of the surface luminosity (in solar units) of the ZAHB at $\log(T_{eff})=3.85$.}
\tablenotetext{f}{Absolute visual magnitude of the ZAHB at $\log(T_{eff})=3.85$.}
}  
\label{zahb}
\end{deluxetable}



\clearpage
       
\clearpage 

\clearpage 

\clearpage

\clearpage

\clearpage

\clearpage 

\clearpage 

\clearpage 

\clearpage   

\clearpage   

\clearpage  

\clearpage  

\clearpage  

\clearpage  

\clearpage  

\clearpage  

\clearpage  

\clearpage  

\clearpage

\clearpage

\clearpage


\begin{thebibliography}{}



\bibitem[]{1024} Ajhar, E.A., \& Tonry, J.L. 1994, ApJ, 429, 557

\bibitem []{a} Alexander, D. R., \&  Ferguson, J. W. 1994, ApJ, 437, 879

\bibitem[]{1028} Bedin, L.R., Cassisi, S., Castelli, F., 
Piotto, G., Anderson, J., Salaris, M., Momany, Y., \& Pietrinferni, A.,
2005, MNRAS, 357, 1038

\bibitem[]{1032} Bellazzini, M., Fusi Pecci, F., Ferraro, F.R., 
Galleti, S., Catelan, M., \& Landsman, W.B. 2001, AJ 122, 2569

\bibitem []{} Bencivenni, D., Castellani, V., Tornamb\'e, A., \& Weiss, A. 1989, ApJS, 71, 109

\bibitem[]{1037} Bergbusch, P.A., \& VandenBerg, D.A. 1992, ApJS, 81, 163

\bibitem[]{1039} B\"ohm-Vitense, E. 1958, Z. Astrophys., 46, 108

\bibitem[]{1041} Brocato, E., Castellani, V., Poli, F.M., \&  Raimondo,
G.2000, A\&AS, 146, 91

\bibitem[]{1044} Cantiello, M., Raimondo, G., Brocato, E., \& Capaccioli,
M. 2003, AJ, 125, 2783

\bibitem []{} Caputo, F., Castellani, V., Chieffi, A., Pulone, L., \& 
Tornamb\'e, A. 1989, ApJ, 340, 241

\bibitem[]{1050} Carney, B.W. 1996, PASP, 108, 900

\bibitem []{w} Carpenter, J. M. 2001, AJ, 121, 2851

\bibitem[]{1054} Carretta, E., \& Gratton, R.G. 1997, A\&AS, 121, 95

\bibitem[]{1056} Carretta, E., Gratton, R.G. Clementini, G., Fusi Pecci,
F. 2000, ApJ, 533, 215

\bibitem[]{1059} Cassisi, S., Salaris, M., Castelli, F., \& Pietrinferni,
A. 2004, ApJ, 616, 498

\bibitem[]{1062} Castelli, F., \& Kurucz, R.L. 2003, IAU  Symposium 210, eds. N. Piskunov, 
W.W. Weiss, and D.F. Gray, p.A20 (astro-ph/0405087) 

\bibitem[]{1065} Chaboyer, B., Sarajedini, A., \& Demarque, P. 1992, ApJ,
394, 515

\bibitem[]{1068} Cox, J.P., \& Giuli, R.T. 1968, Principles of stellar
structure (London: Gordon \& Breach)

\bibitem[]{1071}De Angeli, F., Piotto, G., Cassisi, S., Busso, G., Recio-Blanco, A.,
Salaris, M., Aparicio, A., \& Rosenberg, A. 2005, AJ, 130, 116

\bibitem[]{1074} Del Principe, M., Piersimoni, A.M., Bono, G., Di Paola,
A., Dolci, M., \& Marconi, M. 2005, AJ, 129, 2714 

\bibitem[]{1077} Dorman, B. 1992, ApJS, 81, 221

\bibitem []{} Girardi, L., Bressan, A., Bertelli, G., \& Chiosi, C. 2000,
A\&AS, 141, 371

\bibitem[]{1082} Gratton,R., Sneden, C., \& Carretta, E. 2004, ARAA, 42, 385

\bibitem []{} Grevesse, N., \& Noels, A. 1993, in Origin and Evolution of the
Elements, ed. N. Prantzos, E. Vangioni-Flam, \& M. Cass\'e (Cambridge: Cambridge Univ. Press), 14 

\bibitem[]{1087} Grundahl, F., VandenBerg, D.A., \& Andersen,
M.I. 1998, ApJ, 179, L182

\bibitem[]{1090} Grundahl, F., VandenBerg, D.A., Bell, R.A., Andersen,
M.I., \& Stetson, P.B. 2000, AJ, 120, 1884

\bibitem[]{1093} Harris, W.E. 1996, AJ, 112, 1487

\bibitem[]{1095} Kim., Y.-C., Demarque, P., Yi, S., \& Alexander,
D.R. 2002, ApJS, 143, 499

\bibitem[]{1098} Kraft, R.P., \& Ivans, I.I. 2003, PASP, 115, 143

\bibitem []{at} Iglesias, C. A., \& Rogers, F. J. 1996, ApJ, 464, 943

\bibitem []{} Marigo, P., Bressan, A., \& Chiosi, C. 1996, A\&A, 313, 545

\bibitem[]{1104} McWilliam, A., \& Rich, M.R. 2004, in Origin and
evolution of the elements, A. McWilliam and M. Rauch eds., Carnegie
Observatories Astrophysics Series, vol.4, p.38

\bibitem[]{1108} Pietrinferni, A., Cassisi, S., Salaris, M., \& Castelli, F. 2004, ApJ, 612, 168

\bibitem []{} Potekhin, A. Y., Chabrier, G., \& Shibanov, Y. A. 1999, PhRvE, 60, 2193

\bibitem[]{1112} Recio-Blanco, A. et al. 2005, A\&A, 432, 851


\bibitem[]{1116} Reimers, D. 1975, Mem. Soc. R. Sci. Li\'ege, 8, 369

\bibitem[]{1118} Relyea, L. J., \& Kurucz, R. L. 1978, ApJS, 37, 45 

\bibitem[]{1120} Ryan, S., Norris, J.E., \& Bessel, M.S. 1991, AJ, 102, 303

\bibitem[]{1122} Salaris, M., Chieffi, A., \& Straniero, O. 1993, ApJ,
414, 580

\bibitem []{} Salaris, M., Degl'Innocenti, S., \& Weiss, A. 1997, ApJ, 484, 986

\bibitem []{} Salaris, M., \& Weiss, A. 1998, A\&A, 335, 943

\bibitem[]{1129} Salaris, M., Riello, M., Cassisi, S., \& Piotto, G. 2004,
A\&A, 420, 911

\bibitem[]{1132} Salasnich, B., Girardi, L., Weiss, A., \& Chiosi,
C. 2000, A\&A, 361, 1023 

\bibitem[]{1135} Schlegel, D.J., Finkbeiner, D.P., \& Davis, M. 1998, ApJ,
500, 525

\bibitem[]{1138} Sneden, C. 2004, Mem.S.A.It., 75, 267

\bibitem []{cc} Spergel, D. N., Verde, L., Peiris, H. V., et al. 2003, 
ApJS, 148, 175

\bibitem[]{1143} Tantalo, R., Chiosi, C., \& Bressan, A. 1998, A\&A, 333, 419

\bibitem[]{1145} Trager, S.C., Faber, S.M., Worthey, G., \& Gonzalez,
J.J. 2000, AJ, 119, 1645

\bibitem[]{1148} Valenti, E., Origlia, L., \& Ferraro, F.R. 2005, MNRAS,
361, 272

\bibitem []{ch} VandenBerg, D. A. 2000. ApJS, 129, 315

\bibitem[]{1153} VandenBerg, D.A., \& Irwin, A.W. 1997, in Advances in
Stellar Evolution, Cambridge University Press, p.22

\bibitem []{} VandenBerg, D. A., Swenson, F. J., Rogers, F. J., Iglesias, C. A., 
\& Alexander, D. R. 2000, ApJ, 532, 430

\bibitem[]{1159} Zinn, R., \& West, M.J. 1984, ApJS, 55, 45 

\bibitem[]{1161} Walker, A.R. 1994, AJ, 108, 555

\bibitem[]{1163} Weiss, A., Peletier, R.F., \& Matteucci, F. 1995, A\&A,
296, 73

\bibitem[]{1166} Worthey, G., Faber, S.M., \& Gonzalez, J.J. 1992, ApJ, 398, 69

\bibitem []{ck} Yi, S., Demarque, P., Kim, Y.-C., Lee, Y.-W., Ree, C. H., Lejeune, T.,
\& Barnes, S., 2001, ApJ, 136, 417

\end{thebibliography}
\end{document}